\documentclass[12pt,twoside,a4paper]{article}
\pdfoutput=1

\usepackage{amsmath, amsthm, amsfonts, amssymb}
\usepackage{graphicx}
\usepackage{multirow}
\usepackage{float}
\usepackage{color}
\usepackage{cite}
\usepackage[width=\textwidth,font=small,labelfont=bf,
labelsep=endash]{caption}
\usepackage{titlesec}

\titleformat{\section}
  {\normalfont\fontsize{13}{16}\bfseries}{\thesection}{1em}{}
\titleformat{\subsection}
  {\normalfont\fontsize{11}{14}\bfseries}{\thesubsection}{1em}{}

\normalsize
\newcommand\Poincare          {Poincar\'e\ }

\renewcommand{\theequation}{\arabic{section}.\arabic{equation}}
\makeatletter
\newcommand{\vast}{\bBigg@{4}}
\newcommand{\Vast}{\bBigg@{5}}
\makeatother

\begin{document}

\title{\textbf{Geometric Flow Description of Minimal Surfaces}}
\author{Dimitrios Katsinis$^{1,2}$, Ioannis Mitsoulas$^2$ and Georgios Pastras$^2$}
\date{\small $^1$Department of Physics, National and Kapodistrian University of Athens,\\University Campus, Zografou, Athens 15784, Greece\\
$^2$NCSR ``Demokritos'', Institute of Nuclear and Particle Physics,\\Aghia Paraskevi 15310, Attiki, Greece\linebreak \vspace{8pt}
\texttt{dkatsinis@phys.uoa.gr, mitsoulas@inp.demokritos.gr, pastras@inp.demokritos.gr}}

\vskip .5cm

\maketitle

\begin{abstract}
We introduce a description of a minimal surface in a space with boundary, as the world-hypersurface that the entangling surface traces. It does so by evolving from the boundary to the interior of the bulk under an appropriate geometric flow, whose parameter is the holographic coordinate. We specify this geometric flow for arbitrary bulk geometry. In the case of pure AdS spaces, we implement a perturbative approach for the solution of the flow equation around the boundary. We systematically study both the form of the perturbative solution as well as its dependence on the boundary conditions. This expansion is sufficient for the determination of all the divergent terms of the holographic entanglement entropy, including the logarithmic universal terms in odd spacetime bulk dimensions, for an arbitrary entangling surface, in terms of the extrinsic geometry of the latter.
\end{abstract}

\newpage

\tableofcontents

\newpage

%%%-----------------------------------------------------------------------------------------------------------------------------------------------------------------
\setcounter{equation}{0}
\section{Introduction}
\label{sec:intro}

The holographic duality \cite{Maldacena:1997re,Gubser:1998bc,Witten:1998qj} is a broad framework that connects gravitational theories in spacetimes with AdS asymptotics to conformal field theories on the AdS boundary. As a weak to strong duality it has opened up many new directions for the study of strongly coupled conformal field theories through their weakly coupled gravitational duals.

An important entry in the holographic dictionary was introduced by Ryu and Takayanagi \cite{Ryu:2006bv,Ryu:2006ef}. This establishes a connection between the entanglement entropy in the boundary theory and the area of minimal surfaces in the bulk. More specifically, assuming that the boundary is divided into two subsystems $A$ and $A^C$ by the entangling surface $\partial A$, the entanglement entropy corresponding to this separation of the degrees of freedom is proportional to the area of the open co-dimension two minimal surface in the bulk, which is anchored at the entangling surface, namely
\begin{equation}
{S_{\mathrm{EE}}} = \frac{1}{{4{G_N}}}\mathrm{Area}\left( {{A^{extr}}} \right) .
\end{equation}

The entanglement entropy is a widely used measure of quantum entanglement. It has been shown that it plays an important role in various quantum phenomena (e.g. it is an order parameter in quantum phase transitions \cite{Vidal:2002rm}). In field theory, the calculation of the entanglement entropy is a task that presents many difficulties. Most calculations (see e.g. \cite{Calabrese:2004eu}) incorporate the so called ``replica trick'' \cite{Callan:1994py}. The Ryu-Takayanagi formula has provided the tools for the study of such phenomena through the machinery of the holographic duality, thus in strongly coupled conformal field theories, which are extremely difficult to be studied directly.

In general, the holographic entanglement entropy is divergent. Considering the case of AdS$_{d + 1}$ spacetime and introducing a UV radial cut-off $\Lambda$, it has an expansion of the form \cite{Ryu:2006bv,Ryu:2006ef,Nishioka:2009un}
\begin{equation}
{S_{\mathrm{EE}}} = \begin{cases}
a_{d - 2} \Lambda^{d - 2} + a_{d - 4} \Lambda^{d - 4} + \cdots + a_0 \ln \Lambda / R + \textrm{ regular terms} , & d \textrm{ even}, \\
a_{d - 2} \Lambda^{d - 2} + a_{d - 4} \Lambda^{d - 4} + \cdots + a_0 + \textrm{ regular terms} , & d \textrm{ odd}.
\end{cases}
\end{equation}
The most divergent term is proportional to the area of the entangling surface. This is in agreement with older studies that indicate that the entanglement entropy in (not necessarily conformal) field theory is dominated by  an ``area law'' term \cite{Bombelli:1986rw,Srednicki:1993im}. This is an intriguing similarity to the black hole entropy, which has initiated a large discussion in the literature about whether the black hole entropy can be attributed, totally or partially, to entanglement entropy \cite{Solodukhin:2011gn} and about whether gravity itself can be described as an entropic force due to quantum entanglement statistics \cite{VanRaamsdonk:2009ar,VanRaamsdonk:2010pw}.

The study of the holographic entanglement entropy for arbitrary entangling surfaces is motivated by the underlying relation of the latter to the central charges of the dual CFT. The coefficient of the logarithmic term for even $d$ is universal (i.e. it does not depend on the regularization scheme). This coefficient depends on the values of the central charges of the dual CFT. Since these are related to the holographic Weyl anomaly \cite{Henningson:1998gx}, they can be calculated independently. The consistency of all the relevant calculations is a highly non-trivial check of the holographic duality.

In general the divergent terms of the holographic entanglement entropy, including the universal logarithmic terms, depend on the geometric characteristics of the entangling surface, such as its curvature. In \cite{Solodukhin:2008dh}, the logarithmic term in the case $d = 4$ was connected to the extrinsic geometry of the entangling surface. It was shown to be proportional to the integral of the square of the mean curvature over the whole entangling surface.

A great difficulty that appears in the study of the holographic entanglement entropy is the lack of explicitly known, non-trivial minimal surfaces. Most of the literature focuses on simple cases, like the minimal surfaces that correspond to spherical entangling surfaces. In this work, we use a systematic perturbative approach for the study of minimal surfaces for arbitrary boundary conditions \cite{Schwimmer:2008yh}. We incorporate a description of the minimal surface as the world-hypersurface that the entangling surface traces, as it evolves from the boundary to the interior of the bulk under an appropriate geometric flow, whose parameter is the holographic coordinate. We cast this geometric flow in the form of a simple equation and study in detail its perturbative solution. This is a second order equation, thus its solution depends on both Dirichlet and Neumann boundary conditions. The divergent terms of the holographic entanglement entropy (including the universal logarithmic terms for even $d$) can be specified by this perturbative solution and they depend solely on the Dirichlet boundary data. 

The structure of this paper is as follows: In section \ref{sec:flow} we derive the equations that describe the minimal surface in a space with boundary as a flow of the entangling surface towards the interior of the space. In section \ref{sec:solution} we solve perturbatively the flow equation around the boundary in the case of pure AdS. In section \ref{sec:entropy}, based on the perturbative solution of the previous section, we calculate the divergent terms of the area of the minimal surface. In section \ref{sec:discussion} we discuss our results and possible extensions. Finally, there are some appendices; in appendix \ref{sec:flow_derivation}, we provide some more technical details on the derivation of the flow equation, in appendix \ref{sec:simple solutions}, we show that an explicitly known non-trivial minimal surface in pure AdS$_4$, namely the helicoid, satisfies the flow equation and in appendix \ref{sec:terms_simple} we calculate all divergent terms of the minimal surface area in the case of a spherical entangling surface in order to be used as a verification check for the results of section \ref{sec:entropy}.

\setcounter{equation}{0}
\section{Geometric Flow Description for Minimal Surfaces}
\label{sec:flow}

We desire to describe the minimal surface as a geometric flow, whose parameter is the holographic coordinate. As we move from the boundary towards the interior of the bulk, the entangling surface must evolve under this flow in such a way that it traces the minimal surface. For this purpose, we need to parametrize the minimal surface appropriately; one of the parameters should be identical to the value of the holographic coordinate. Furthermore, we need to study the intersections of the minimal surface with the planes where the holographic coordinate is constant. Subsequently, we will specify, how these intersections must evolve as the holographic coordinate changes, so that their union is the minimal surface.

\subsection{Background Geometry}
\label{subsec:flow_background}

We herein focus our attention on \emph{static}, asymptotically AdS$_{d+1}$ spacetimes, although our analysis applies to any static spacetime with a boundary. We further demand that the entangling surface is time-independent. It follows that the co-dimension two minimal surface that is involved in the Ryu-Takayanagi formula is also time-independent. Therefore, the problem of its specification can be reduced to one of finding a co-dimension one minimal surface in an asymptotically hyperboloid Riemannian space, which is a time-slice of the original spacetime.

In the following, $r$ denotes the holographic coordinate and $x^i$, $i = 1 , \cdots , d-1$, denote the rest of the coordinates. Furthermore, we select a coordinate system so that the metric of the asymptotically hyperboloid Riemannian space assumes the form
\begin{equation}
d{s^2} = f\left( r \right)d{r^2} + {h_{ij}}\left( {r,{x^k}} \right)d{x^i}d{x^j} .
\label{eq:metric_G}
\end{equation}
The metric can always be written in such a form via an appropriate redefinition of the holographic coordinate $r$. The space boundary in these coordinates is described by an equation of the form $r = r_0$ (e.g. in the case of pure AdS, in \Poincare coordinates $r_0 = 0$, whereas in global coordinates $r_0 = \infty$).

We also consider the constant-$r$ slices of this space. On the slice $r = \rho$, the induced metric is given by
\begin{equation}
d{s^2} = {h_{ij}}\left( {\rho;{x^k}} \right)d{x^i}d{x^j} .
\label{eq:metric_g}
\end{equation}

Using the form of the metric \eqref{eq:metric_G}, we can calculate the Christoffel symbols
\begin{equation}
\begin{split}
&\Gamma _{rr}^r = \frac{1}{2}\frac{{f'\left( r \right)}}{{f\left( r \right)}} , \quad \Gamma _{ri}^r = 0 , \quad \Gamma _{ij}^r =  - \frac{1}{2}\frac{{{\partial _r}{h_{ij}}}}{{f\left( r \right)}}\\
&\Gamma _{rr}^i = 0 , \quad \Gamma _{rj}^i = \frac{1}{2}{h^{ik}}{\partial _r}{h_{kj}} , \quad \Gamma _{jk}^i = \gamma _{jk}^i ,
\end{split}
\label{eq:Christoffel_G}
\end{equation}
where $ \gamma _{jk}^i$ are the Christoffel symbols with respect to the induced metric on the constant-$r$ slices \eqref{eq:metric_g}. In the following, the capital letters refer to quantities defined in the bulk and the corresponding lowercase ones refer to the corresponding quantities defined in the constant-$r$ slices.

\subsection{Two Embedding Problems}
\label{subsec:flow_embeddings}
\begin{figure}[ht]
\vspace{10pt}
\begin{center}
\begin{picture}(60,35)
\put(5,0){\includegraphics[width = 0.6\textwidth]{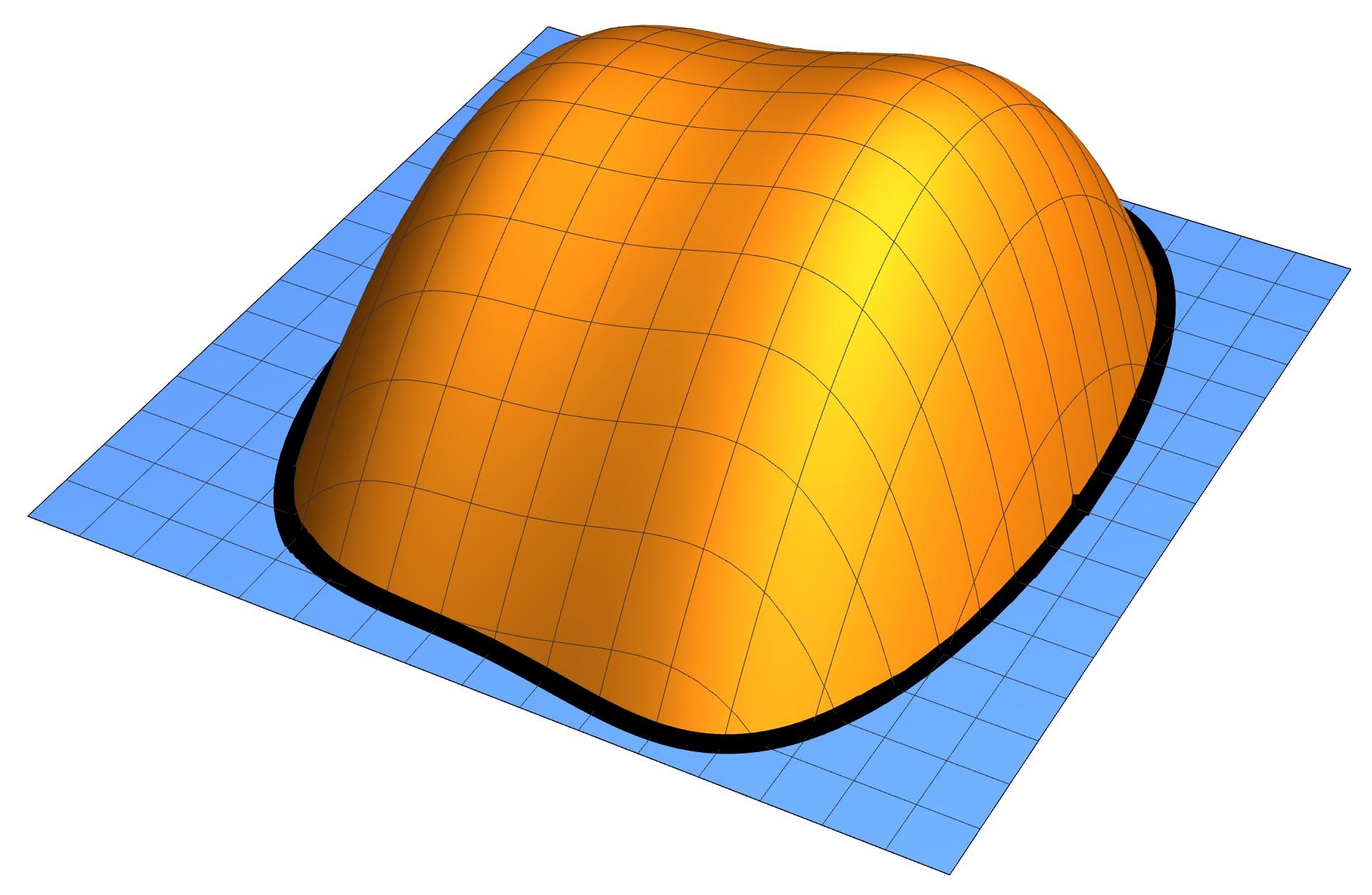}}
\put(30,26.5){\rotatebox{-7}{$A_{\textrm{minimal}}$}}
\put(47,6){\rotatebox{57}{AdS boundary}}
\put(21,14.5){\rotatebox{-25}{$C_{\textrm{entangling}}$}}
\end{picture}
\end{center}
\vspace{-10pt}
\caption{The embedding of the minimal surface in the asymptotically hyperbolic space}
\vspace{5pt}
\label{fig:embedding1}
\end{figure}
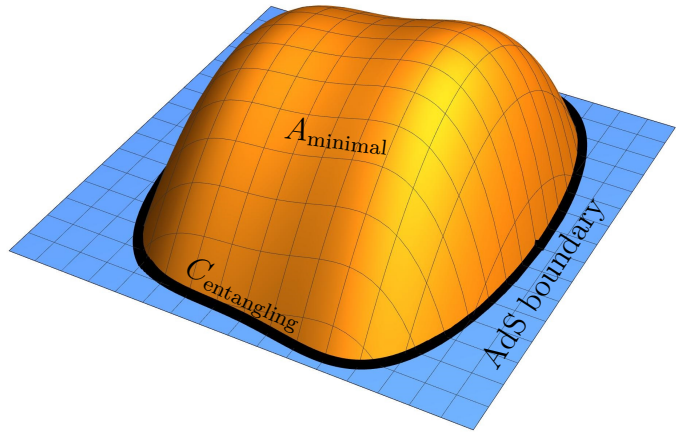

We consider two embedding problems. The first one is the embedding of the minimal surface in the asymptotically hyperboloid space, which is depicted in figure \ref{fig:embedding1}. The minimal surface is parametrized by $\rho$ and $u^a$, where $a = 1 , \cdots , d-2$, so that
\begin{equation}
\begin{split}
r &= \rho , \\
{x^i} &= {X^i}\left( {\rho,{u^a}} \right) ,
\end{split}
\label{eq:parametrization_bulk}
\end{equation}
i.e., one of the parameters \emph{equals the value of the holographic coordinate} $r$. In the following, the indices $i$, $j$ and so on, refer to the coordinates on a constant-$r$ plane and take values from 1 to $d - 1$, whereas the indices $a$, $b$ and so on, refer to the parameters $u^a$ and take values from 1 to $d - 2$.

\begin{figure}[ht]
\vspace{10pt}
\begin{center}
\begin{picture}(100,37)
\put(0,0){\includegraphics[width = 0.5\textwidth]{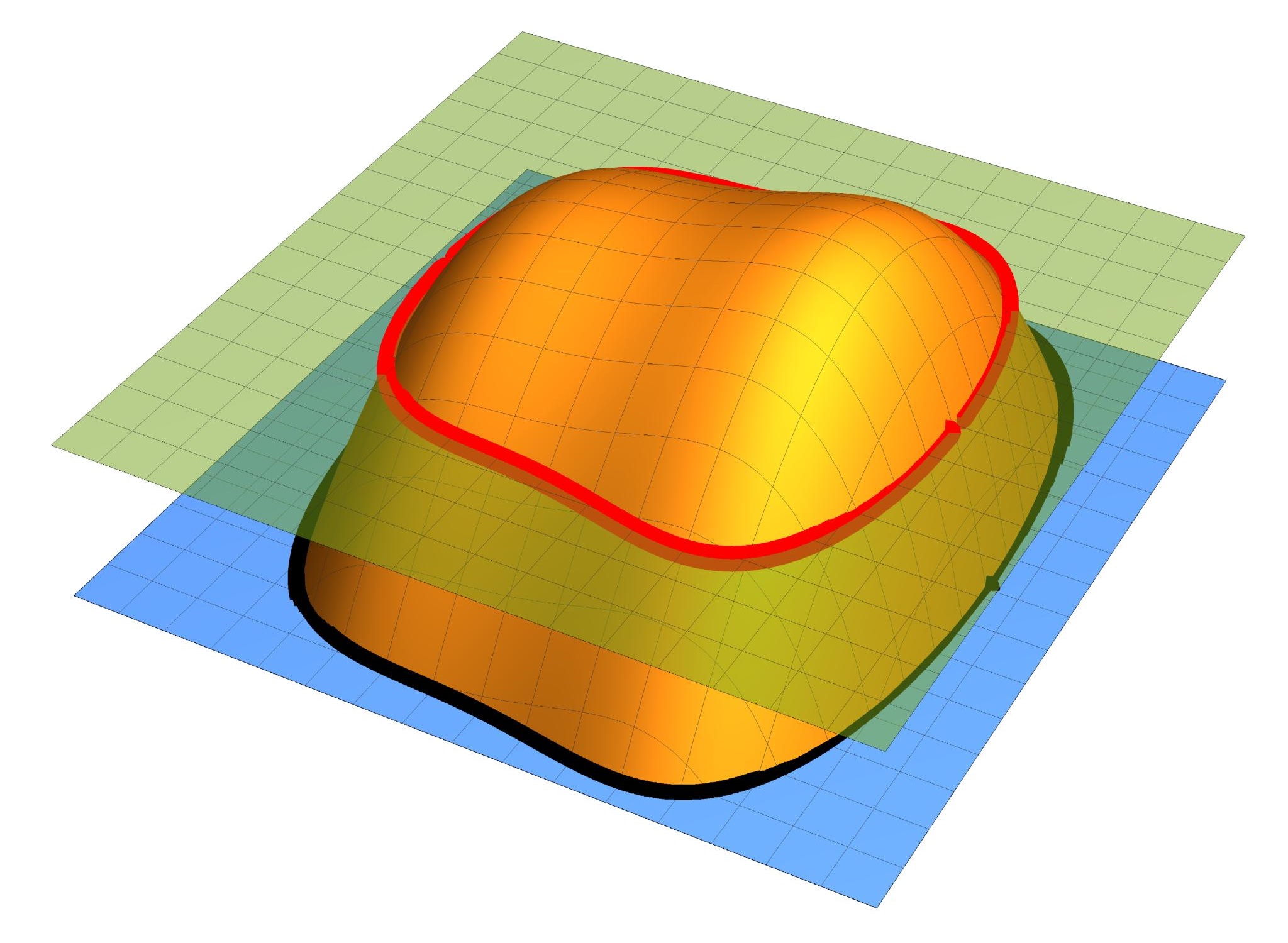}}
\put(50,0){\includegraphics[width = 0.5\textwidth]{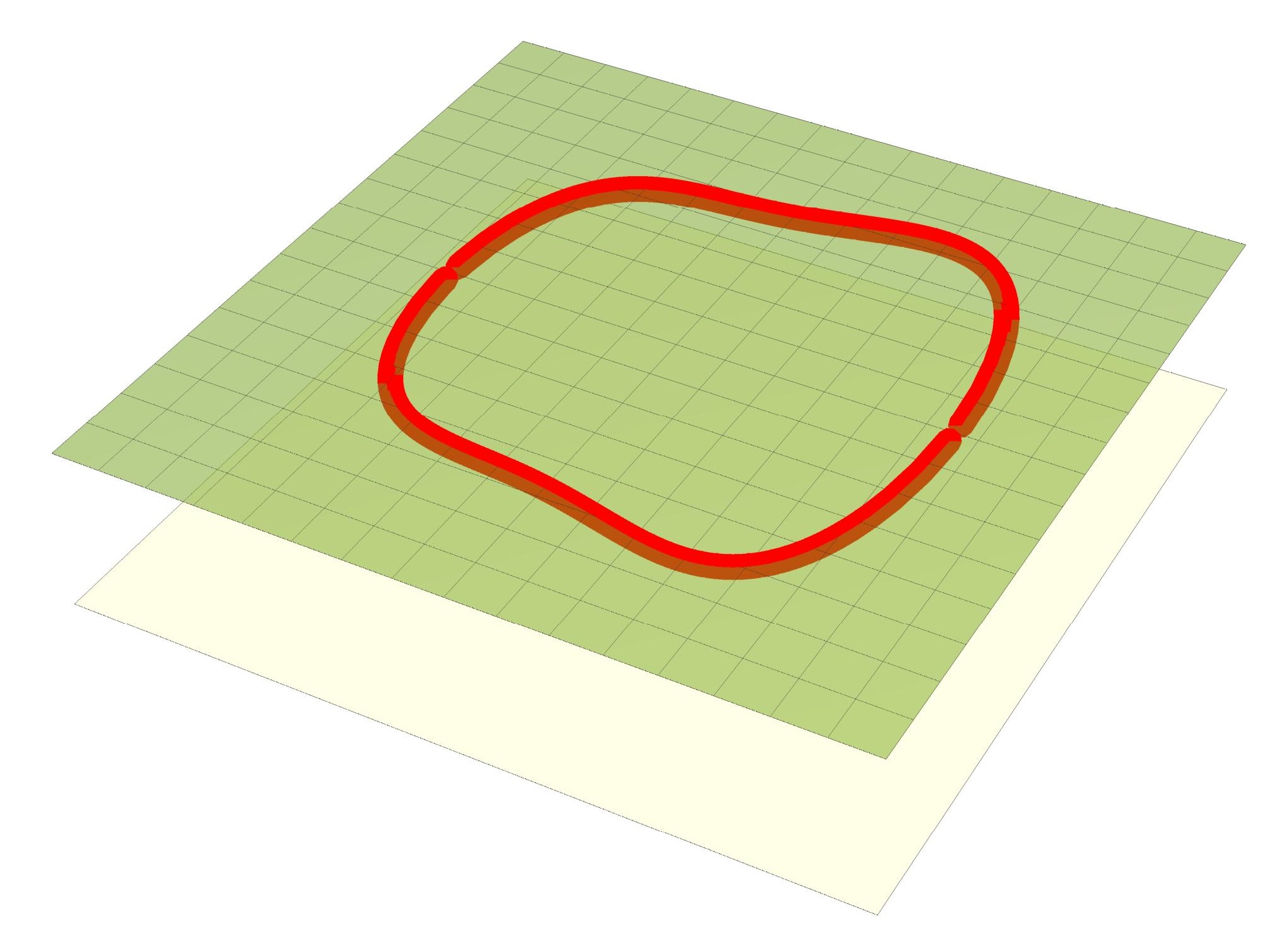}}
\put(21,24){\rotatebox{-7}{$A_{\textrm{minimal}}$}}
\put(33.5,4){\rotatebox{57}{AdS boundary}}
\put(13,12.5){\rotatebox{-23}{$C_{\textrm{entangling}}$}}
\put(7,21.25){\rotatebox{42}{$r=\rho$ plane}}
\put(57,21.25){\rotatebox{42}{$r=\rho$ plane}}
\put(74,30){\rotatebox{-9}{Intersection}}
\put(72.75,26.75){\rotatebox{-9}{with $A_{\textrm{minimal}}$}}
\end{picture}
\end{center}
\vspace{-10pt}
\caption{On the left, the intersection of the minimal surface with a constant-$r$ plane. On the right, the embedding of the intersection in the constant-$r$ plane.}
\vspace{5pt}
\label{fig:embedding2}
\end{figure}
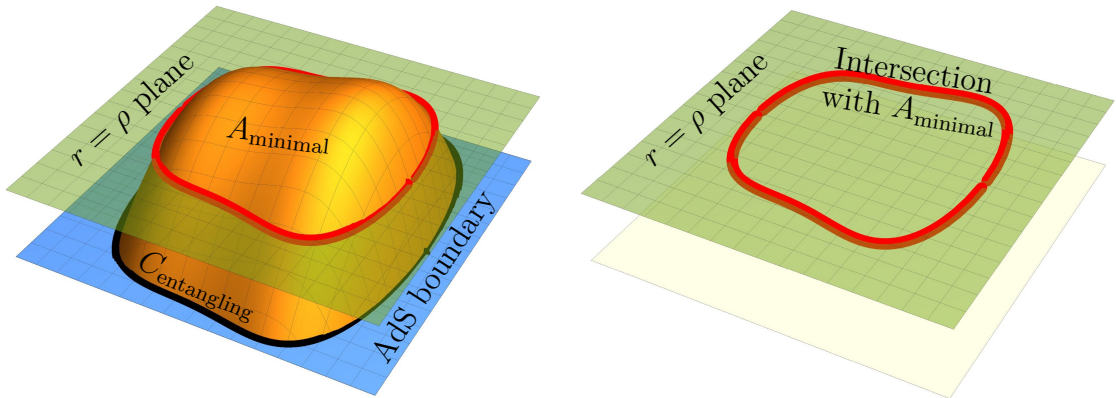

Similarly, we consider the embedding of the intersection of the minimal surface with a constant-$r$ plane in this constant-$r$ plane, as shown in figure \ref{fig:embedding2}.
Assuming that the latter is described by the equation $r=\rho$, we parametrize the aforementioned intersection as
\begin{equation}
{x^i} = {x^i}\left( {{\rho};{u^a}} \right) ,
\end{equation}
where ${x^i}\left( {\rho;{u^a}} \right) = {X^i}\left( {\rho,{u^a}} \right)$. The functions ${X^i}\left( {\rho,{u^a}} \right)$ should be considered as functions of $d-1$ coordinates, whereas the functions ${x^i}\left( {{\rho};{u^a}} \right)$ should be considered as functions of $d-2$ coordinates and a parameter $\rho$, identifying the constant-$r$ plane. Obviously, at the limit $\rho \to r_0$ the intersection of the minimal surface with a constant-$r$ plane tends to the intersection of the minimal surface with the boundary, i.e. the entangling surface. Since the functions ${X^i}\left( {\rho,{u^a}} \right)$ and ${x^i}\left( {\rho;{u^a}} \right)$ are identical, we will avoid using both symbols in the following. Our goal is to express the minimal surface as a flow of the entangling surface towards the interior of the bulk. For this reason, we choose to use the lowercase notation ${x^i}\left( {\rho;{u^a}} \right)$ and we will drop its arguments in what follows. Similarly, we will drop the arguments of the induced metric $h$, keeping in mind that it depends on the parameter $\rho$ both explicitly and implicitly, as it takes values on the intersection with the minimal surface. The explicit derivative will be denoted by $\partial_r h_{ij}$, whereas the total derivative with respect to parameter $\rho$ will be denoted by $\partial_\rho h_{ij}$, i.e. $\partial_\rho h_{ij} = \partial_r h_{ij} + \frac{{\partial {x^k}}}{{\partial \rho}} \partial_k h_{ij}$.

We adopt the notation $A^\mu = \left( A^r , A^i \right)$ for vectors in the bulk. We define the following $d - 1$ vectors, which are tangent to the minimal surface 
\begin{equation}
T_\rho^\mu  = \left( {1,\frac{{\partial {x^i}}}{{\partial \rho}}} \right) , \quad T_a^\mu  = \left( {0,\frac{{\partial {x^i}}}{{\partial {u^a}}}} \right) .
\label{eq:embedding_T}
\end{equation}
We also have $d - 2$ vectors in the $r = \rho$ plane, which are tangent to the intersection of the minimal surface with the plane. These are
\begin{equation}
t_a^i \left( {\rho} \right) = \frac{{\partial {x^i}}}{{\partial {u^a}}} .
\label{eq:embedding_t}
\end{equation}

Both embedding problems are co-dimension one problems, thus, in both cases there is a single normal vector. Let the normal vector of the bulk problem be $N$. Then, it obeys
\begin{align}
{N^r}f\left( \rho \right) + {N^i}\frac{{\partial {x^j}}}{{\partial \rho}}{h_{ij}} &= 0 , \label{eq:embedding_N_per_rho}\\
{N^i}\frac{{\partial {x^j}}}{{\partial {u^a}}}{h_{ij}} &= 0 . \label{eq:embedding_N_per_a}
\end{align}
Furthermore, demanding that the normal vector is normalized implies that
\begin{equation}
{\left( {{N^r}} \right)^2}f\left( \rho \right) + {N^i}{N^j}{h_{ij}} = 1 .
\label{eq:embedding_N_norm}
\end{equation}
Similarly, the normal vector $n$ in the constant-$r$ plane must obey
\begin{equation}
{n^i}\frac{{\partial {x^j}}}{{\partial {u^a}}}{h_{ij}} = 0 ,
\label{eq:embedding_n_per}
\end{equation}
so that it is perpendicular to the tangent vectors $t_a$ and
\begin{equation}
{n^i}{n^j}{h_{ij}} = 1,
\label{eq:embedding_n_norm}
\end{equation}
so that it is normalized.

The equations \eqref{eq:embedding_N_per_a} and \eqref{eq:embedding_n_per} imply that at a given $r = \rho$ plane, the normal vector $n$ and the projection of the normal vector $N$ on this plane are parallel, i.e.
\begin{equation}
{N^i} = c\left( {\rho;{u^a}} \right){n^i} .
\end{equation}
Furthermore, the equation \eqref{eq:embedding_N_per_rho} implies that
\begin{equation}
{N^r} =  - \frac{1}{{f\left( \rho \right)}}{N^i}\frac{{\partial {x^j}}}{{\partial \rho}}{h_{ij}} =  - \frac{{c\left( {\rho;{u^a}} \right)}}{{f\left( \rho \right)}}{n^i}\frac{{\partial {x^j}}}{{\partial \rho}}{h_{ij}} .
\end{equation}
Finally, the normalization of $N$ \eqref{eq:embedding_N_norm} restricts $c\left( {\rho;{u^a}} \right)$ to be equal to
\begin{equation}
c\left( {\rho;{u^a}} \right) = {\left[ {\frac{1}{{f\left( \rho \right)}}{{\left( {{n^i}\frac{{\partial {x^j}}}{{\partial \rho}}{h_{ij}}} \right)}^2} + 1} \right]^{ - \frac{1}{2}}} .
\label{eq:embedding_c_general_parametrization}
\end{equation}

In the following, we will adopt a specific parametrization of the minimal surface, which simplifies the algebra significantly. As the holographic coordinate $r$ runs, the trace of the minimal surface varies. At a given $r = \rho$ plane, this variation is described by the vector ${\frac{{\partial {x^i}}}{{\partial \rho}}}$. However, any component of this vector that is parallel to the intersection of the minimal surface with the plane corresponds to a reparametrization of the intersection and not to a physical alteration of the latter. As a clarifying example, let us consider the special case where the vector ${\frac{{\partial {x^i}}}{{\partial \rho}}}$ is parallel to the intersection everywhere; then, as $\rho$ varies, the intersection is invariant. It follows that an appropriate choice of the parameters $u^a$ at each $r = \rho$ plane (obviously this is a redefinition of $u^a$ that involves $\rho$) can set ${\frac{{\partial {x^i}}}{{\partial \rho}}}$ parallel to $n^i$, i.e.
\begin{equation}
\frac{{\partial {x^i}}}{{\partial \rho}} = a\left( {\rho;{u^a}} \right){n^i} .
\label{eq:parametrization}
\end{equation}
This is always possible through an appropriate Penrose-Brown-Henneaux transformation \cite{Penrose,Brown:1986nw}. This selection partially fixes the diffeomorphisms of the minimal surface parametrizations. There are remaining diffeomorphisms corresponding to redefinitions of the parameters $u^a$ that do not involve the parameter $\rho$. In the following, we will always use such a parametrization for the minimal surface.

As follows from the equation \eqref{eq:embedding_c_general_parametrization}, for this specific parametrization, the normalization factor $c\left( {\rho;{u^a}} \right)$ assumes the form
\begin{equation}
c\left( {\rho;{u^a}} \right) = {\left( {\frac{{a{{\left( {\rho;{u^a}} \right)}^2}}}{{f\left( \rho \right)}} + 1} \right)^{ - \frac{1}{2}}}
\label{eq:embedding_c_specoal_parametrization}
\end{equation}
and the $r$ component of the normal vector $N$ is written as
\begin{equation}
{N^r} =  - \frac{{c\left( {\rho;{u^a}} \right)a\left( {\rho;{u^a}} \right)}}{{f\left( \rho \right)}} .
\end{equation}

Finally, the elements of the induced metric for the embedding of the minimal surface in the asymptotically hyperboloid space are given by,
\begin{equation}
\begin{split}
{\Gamma _{\rho\rho}} &= f\left( \rho \right) + a{\left( {\rho;{u^a}} \right)^2} ,\\
{\Gamma _{\rho a}} &= 0 ,\\
{\Gamma _{ab}} &= {\gamma _{ab}} ,
\end{split}
\label{eq:embedding_Gamma}
\end{equation}
where ${\gamma _{ab}}$ are the elements of the induced metric for the embedding of the intersection of the minimal surface with the $r = \rho$ plane, in the latter, namely
\begin{equation}
{\gamma _{ab}} = \frac{{\partial {x^i}}}{{\partial {u^a}}}\frac{{\partial {x^j}}}{{\partial {u^b}}}{h_{ij}} .
\label{eq:embedding_g}
\end{equation}
In this parametrization, the elements of the inverse induced metric assume the form
\begin{equation}
\begin{split}
{\Gamma ^{\rho\rho}} &= \frac{1}{{f\left( \rho \right) + a{{\left( {\rho ;{u^a}} \right)}^2}}} = \frac{{c{{\left( {\rho ;{u^a}} \right)}^2}}}{{f\left( \rho \right)}},\\
{\Gamma ^{a\rho}} &= 0 ,\\
{\Gamma ^{ab}} &= {\gamma ^{ab}} .
\end{split}
\label{eq:embedding_inverse_G}
\end{equation}
Notice that the symbols $\gamma$ and $\Gamma$ denote the induced metric elements when they have two indices, whereas they denote the Christoffel symbols \eqref{eq:Christoffel_G}, whenever they have three indices.

We proceed to calculate the corresponding second fundamental forms for the two embeddings under consideration. By definition, the second fundamental form for the intersection of the minimal surface with the $r = \rho$ plane is
\begin{equation}
{k_{ab}} = - {\nabla _k}{n^i}\frac{{\partial {x^k}}}{{\partial {u^a}}}\frac{{\partial {x^j}}}{{\partial {u^b}}}{h_{ij}} = - {\partial _a}{n^i}\frac{{\partial {x^j}}}{{\partial {u^b}}}{h_{ij}} - \gamma _{kl}^i{n^l}\frac{{\partial {x^k}}}{{\partial {u^a}}}\frac{{\partial {x^j}}}{{\partial {u^b}}}{h_{ij}}.
\label{eq:embedding_k}
\end{equation}

It is a matter of algebra, which is included in the appendix \ref{sec:flow_derivation}, to show that the elements of the second fundamental form for the embedding of the minimal surface in the bulk are given by
\begin{equation}
\begin{split}
{K_{\rho\rho}} &= \sqrt {f} c{\partial _\rho}\left( {\frac{a}{{\sqrt {f} }}} \right) + \frac{a}{{2c}}{n^i}{n^j}{\partial _r}{h_{ij}} , \\ 
{K_{\rho a}} &= c{\partial _a}a + \frac{1}{{2c}}{n^i}\frac{{\partial {x^j}}}{{\partial {u^a}}}{\partial _r}{h_{ij}} , \\
{K_{ab}} &= c{k_{ab}} + \frac{{ca}}{{2 f}}\frac{{\partial {x^k}}}{{\partial {u^a}}}\frac{{\partial {x^j}}}{{\partial {u^b}}} {\partial _r}{h_{kj}}.
\end{split}
\label{eq:embedding_K}
\end{equation}
Finally, the mean curvature equals
\begin{equation}
K = ck + \frac{{c^3}}{{\sqrt {f} }}{\partial _\rho}\left( {\frac{a}{{\sqrt {f} }}} \right) + \frac{{ca}}{{2 f}} {h^{ij}} {{\partial _r}{h_{ij}}} .
\label{eq:embedding_trace_K}
\end{equation}

\subsection{The Minimal Surface as a Flow of the Entangling Surface Towards the Interior of the Bulk}

Having studied the two embedding problems in section \ref{subsec:flow_embeddings}, it is simple to find an equation that describes the minimal surface as a surface being traced by the entangling surface, which evolves under an appropriate geometric flow, whose parameter is the holographic coordinate. By definition, the minimal surface satisfies the equation
\begin{equation}
K = 0.
\end{equation}
This combined with the equation \eqref{eq:embedding_trace_K} implies
\begin{equation}
\frac{1}{{\sqrt {f} }}{\partial _\rho}\left( {\frac{a}{{\sqrt {f} }}} \right) + \frac{k}{{{c^2}}} + \frac{{a}}{{2 c^2 f}} {h^{ij}} {{\partial _r}{h_{ij}}} = 0.
\end{equation}
Finally, using the equation \eqref{eq:embedding_c_specoal_parametrization} to eliminate $c$, we arrive at
\begin{equation}
\frac{1}{{2 a}}{\partial _\rho}\left( {\frac{{{a^2}}}{{f}} + 1} \right) + \left( {\frac{{{a^2}}}{{f}} + 1} \right) \left( k + \frac{{a}}{{2 f}} {h^{ij}} {{\partial _r}{h_{ij}}} \right) = 0 .
\label{eq:flow}
\end{equation}

Let us now focus our attention on pure AdS$_{d+1}$ or actually on a time slice of it, the hyperboloid H$^d$. In \Poincare coordinates $f \left( r \right) = 1 / r^2$ and $h_{ij} \left( r ; x^i \right) = \delta_{ij} / r^2$. These imply that ${h^{ij}} {{\partial _r}{h_{ij}}} = - 2 \left( d - 1 \right) / r$. Thus, the equation \eqref{eq:flow} assumes a much simpler form,
\begin{equation}
\rho \partial_\rho \left(\rho a\right) + \left(\rho^2 a^2 + 1\right) \left(k - \left(d - 1\right) \rho a\right) = 0
\label{eq:flow_Hd}
\end{equation}
or
\begin{equation}
\rho \partial_\rho \arctan \left( \rho a \right) + k - \left(d - 1\right) \rho a = 0 .
\end{equation}

It can be easily verified that all known minimal surfaces in H$^d$, such as the minimal surfaces that correspond to a spherical or strip region in the boundary, as well as the catenoid and helicoid minimal surfaces in H$^3$, satisfy the equation \eqref{eq:flow_Hd}. The proof for the non-trivial case of the helicoid is included in the appendix \ref{sec:simple solutions}.

In an isotropic background, such as a time slice of the pure AdS spacetime, the bulk coordinates in a local patch can be selected so that $h_{ij} = g \left( r \right) \delta_{ij}$. For such backgrounds and for this selection of the bulk coordinates, all Christoffel symbols $\gamma^i_{jk}$ vanish and thus the second fundamental form for the embedding of the intersection in the constant-$r$ plane assumes the form
\begin{equation}
{k_{ab}} = - \frac{1}{a}\frac{{\partial^2 {x^i}}}{{\partial {u^a} \partial {\rho}}}\frac{{\partial {x^j}}}{{\partial {u^b}}}{h_{ij}}.
\end{equation}
This further implies that the mean curvature can be written as
\begin{equation}
- 2ak = {\gamma ^{ab}}{\partial _\rho }{\gamma _{ab}} - \frac{{\partial _\rho }{g}}{g}{\gamma ^{ab}}{\gamma _{ab}} = \frac{1}{2}\frac{{{\partial _\rho }\det \gamma }}{{\det \gamma }} - \left({d - 2}\right)\frac{{\partial _\rho }{g}}{g}.
\label{eq:flow_mean_curvature}
\end{equation}
This formula allows the re-expression of the equation \eqref{eq:flow} as
\begin{equation}
{\partial _\rho }\left( {c\sqrt {g \det \gamma }} \right) - \frac{{\left( {d - 1} \right)\sqrt {\det \gamma } }}{{c }} {\partial _\rho \sqrt {g}} = 0.
\label{eq:flow_homogenous}
\end{equation}
In the case of the H$^d$ space, $g \left( r \right) = 1 / r^2$, and thus the equation \eqref{eq:flow_homogenous} assumes the form
\begin{equation}
\rho {\partial _\rho }\left( {\frac{{c\sqrt {\det \gamma } }}{{\rho}}} \right) + \frac{{\left( {d - 1} \right)\sqrt {\det \gamma } }}{{c\rho }} = 0 ,
\label{eq:flow_homogenous_Hd}
\end{equation}
which will become handy in next section.

\subsection{A Comment on the Boundary Conditions}

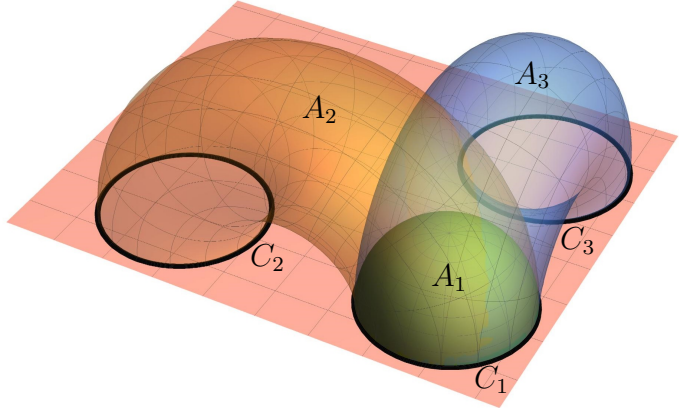
\begin{figure}[ht]
\vspace{10pt}
\begin{center}
\begin{picture}(65,35)
\put(2.5,0){\includegraphics[width = 0.6\textwidth]{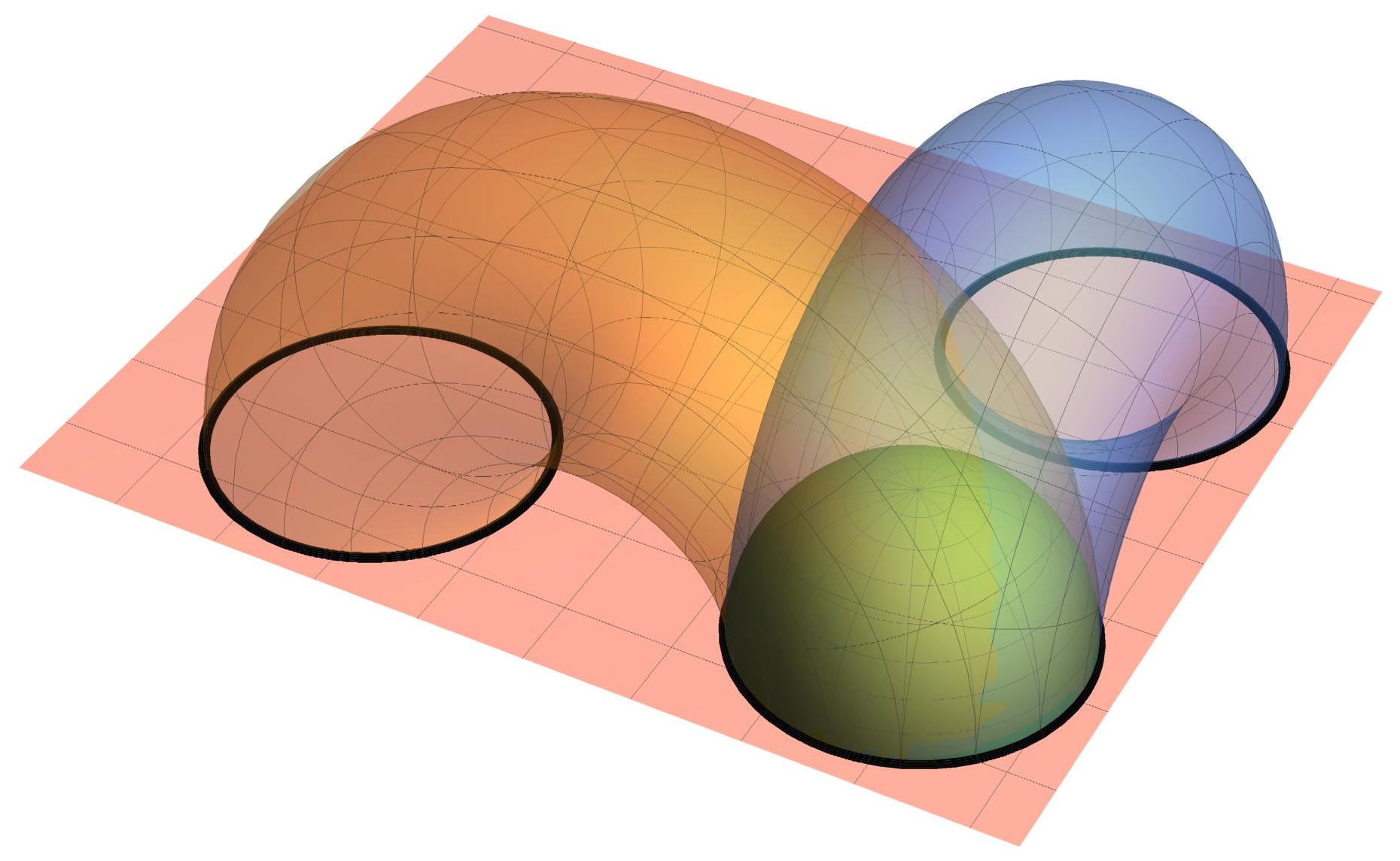}}
\put(40.25,11.25){$A_1$}
\put(29,26){$A_2$}
\put(47.5,29){$A_3$}
\put(44,2.25){$C_1$}
\put(24.5,13){$C_2$}
\put(51.5,14.5){$C_3$}
\end{picture}
\end{center}
\vspace{-10pt}
\caption{Three minimal surfaces. The minimal surface $A_1$ corresponds to the connected entangling surface $C_1$. The minimal surfaces $A_2$ and $A_3$ correspond to the disconnected entangling surfaces $C_1 \cup C_2$ and $C_1 \cup C_3$, respectively.}
\vspace{5pt}
\label{fig:multiple}
\end{figure}

The flow equation \eqref{eq:flow} contains second derivatives of the embedding functions with respect to the holographic coordinate. Therefore, the specification of a connected entangling surface (i.e. a Dirichlet boundary condition), does not uniquely determine the solution of the minimal surface. This is due to the fact that such an entangling surface may be part of a more complex disconnected entangling surface, (see figure \ref{fig:multiple}). The additional Neumann-type boundary condition, which is required for the specification of a unique solution, is equivalent to the specification of the other components of the disconnected entangling surface. Would we desire to find a minimal surface that corresponds to a connected entangling surface, we should specify the additional initial condition in an appropriate fashion. Two clarifying examples that correspond to disconnected entangling surfaces are the minimal surface corresponding to a strip region in H$^d$ and the catenoid surface in H$^3$.

\subsection{A Comment on the Parametrization of the Minimal Surface}
\label{subsec:saddle}

\begin{figure}[ht]
\vspace{10pt}
\begin{center}
\begin{picture}(100,37)
\put(0,0){\includegraphics[width = 0.5\textwidth]{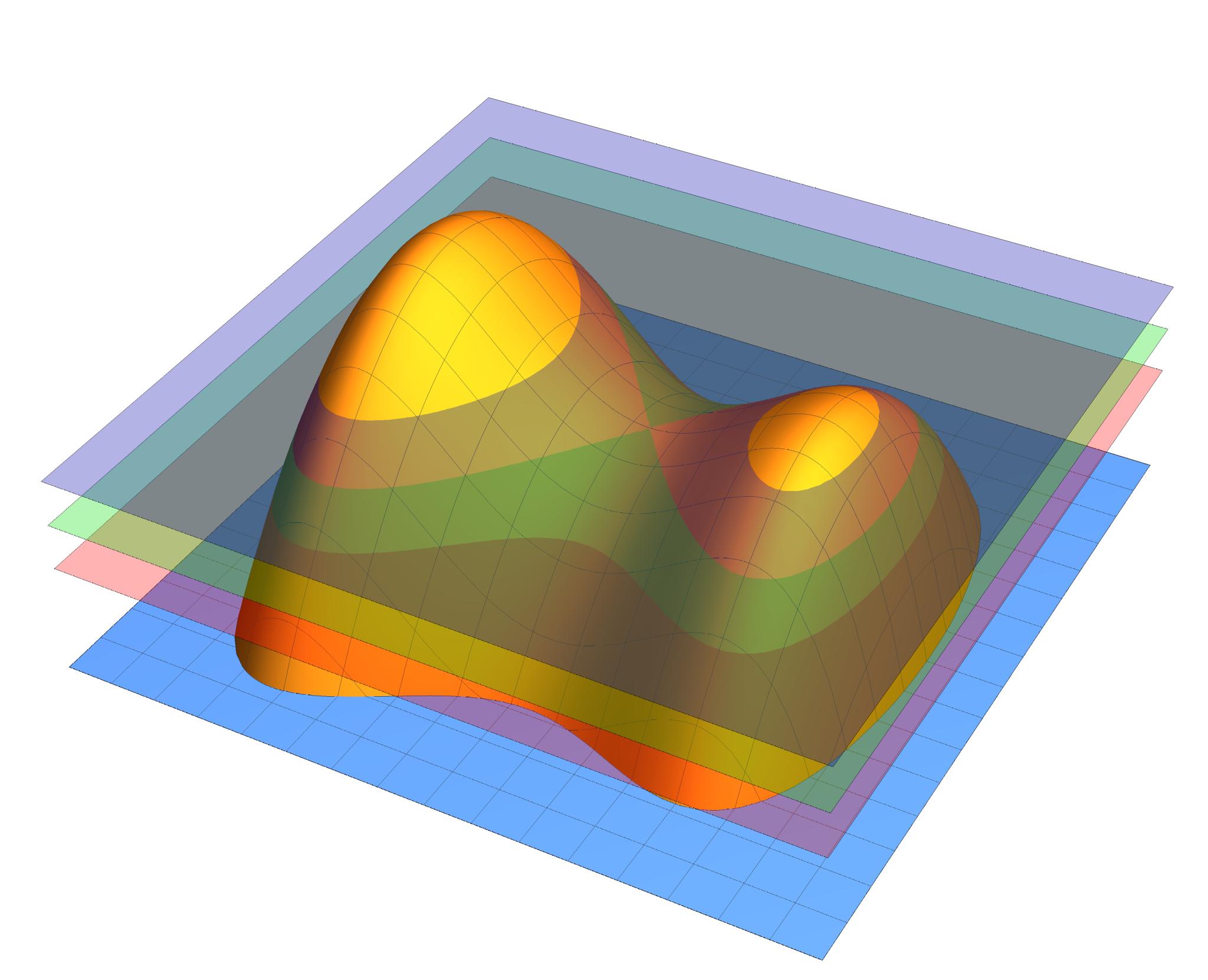}}
\put(50,0){\includegraphics[width = 0.5\textwidth]{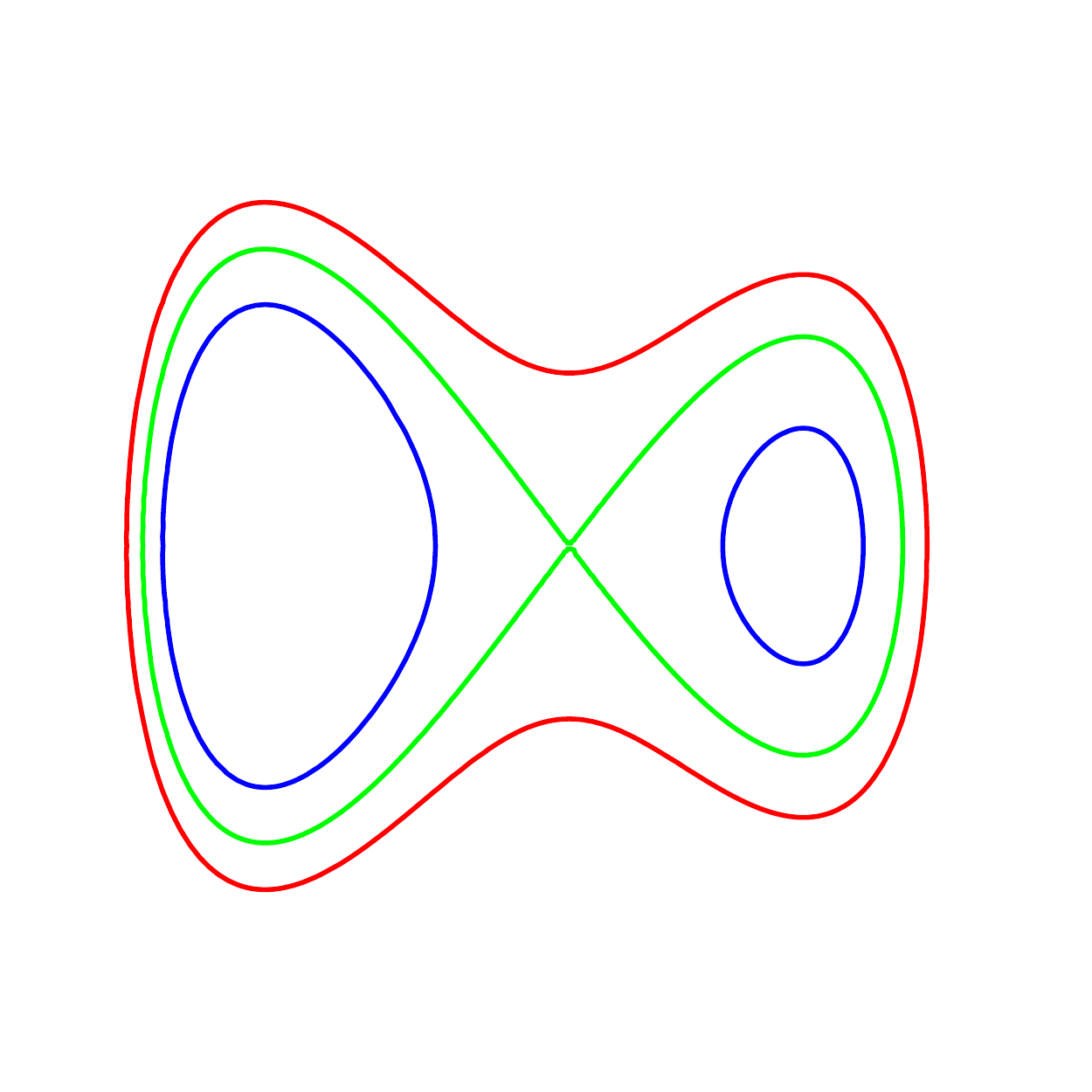}}
\end{picture}
\end{center}
\vspace{-10pt}
\caption{The intersection of the minimal surface with the constant-$r$ planes around a saddle point}
\vspace{5pt}
\label{fig:saddle}
\end{figure}

In the case the minimal surface has a single local maximum of the holographic coordinate, the parametrization \eqref{eq:parametrization} can be applied for the whole minimal surface. This parametrization will have a single singular point, the maximum itself, where the embedding functions will map the whole range of the parameters $u^a$ to the same point. However, if more than one local maxima exist, there is a constant-$r$ plane for a value of the holographic coordinate $r_{\textrm{saddle}}$, smaller than the value of the holographic coordinate at the maxima, which contains a saddle point, as shown in figure \ref{fig:saddle}.
At this constant-$r$ slice, the intersection of the minimal surface is not smooth. At the non-smooth point, the normal vector ceases being well-defined and the definition of the parametrization \eqref{eq:parametrization} becomes problematic. When a saddle point is met, the problem must be split to two new problems whose boundary conditions are defined at $r = r_{\textrm{saddle}}$ in an appropriate fashion, so that the surface is smooth.

The inverse situation occurs in the case of solenoid-like minimal surfaces that correspond to disconnected entangling surfaces in the boundary. In such cases there appear saddle points where two distinct problems merge. At such a saddle point, the demand for the smoothness of the minimal surface will result in constraints to the Neumann conditions that were applied in each of the two separate problems, which in effect will transform each of the two problems, from boundary value problems with one Dirichlet and one Neumann condition to a problem with two Dirichlet conditions.

\setcounter{equation}{0}
\section{The Perturbative Solution to the Flow Equation in Pure AdS$_{d+1}$}
\label{sec:solution}

In this section we will present a perturbative approach, based on the techniques of \cite{Schwimmer:2008yh}, for the solution of the equation \eqref{eq:flow_homogenous_Hd}, which describes the minimal surface as a geometric flow of the entangling surface into the interior of pure AdS space. It has to be noted that a similar approach can be developed for other asymptotically AdS static and isotropic backgrounds on the basis of equation \eqref{eq:flow_homogenous}, or more general static backgrounds on the basis of \eqref{eq:flow}.

\subsection{Set-up of the Perturbative Calculation}

We assume an expansion for the embedding functions of the minimal surface around $\rho=0$ of the form
\begin{equation}
x^i \left( \rho ; u^a \right) = \sum_{m=0}^\infty {x_{\left( m \right)}^i \left( u^a \right) \rho^m}.
\label{eq:perturbation_x_expansion}
\end{equation}
Obviously, the first term in this expansion is determined by the Dirichlet boundary condition, i.e. the entangling surface, which is parametrized by
\begin{equation}
x^i = \mathcal{X}^i \left( u^a \right) = x_{\left( 0 \right)}^i \left( u^a \right) .
\label{eq:perturbation_x_entangling}
\end{equation}

In the following, we will refer to the induced metric and the extrinsic curvature emerging from the embedding functions \eqref{eq:perturbation_x_entangling} and with respect to the metric $\delta_{ij}$, as the induced metric $\mathcal{G}$ and the extrinsic curvature $\mathcal{K}$ of the entangling surface,
\begin{align}
{\mathcal{G}_{ab}} &= {\partial _a}{\mathcal{X}^i}{\partial _b}{\mathcal{X}^i} , \label{eq:perturbation_g_entangling}\\
{\mathcal{K}_{ab}} &=  - {\partial _a}{\mathcal{N}^i}{\partial _b}{\mathcal{X}^i} \label{eq:perturbation_k_entangling},
\end{align}
where ${\mathcal{N}^i}$ is the normal vector of the entangling surface, normalized with respect to the metric $\delta_{ij}$, i.e. ${\mathcal{N}^i} = \mathop {\lim }\limits_{\rho  \to 0} \frac{{{n^i}}}{\rho }$. Here and in the following, the presence of a repeated upper index implies summation over all its values.

It follows that the induced metric $\gamma$ has a similar expansion of the form
\begin{equation}
{\gamma _{ab}} = \frac{1}{{{\rho ^2}}}\sum\limits_{m = 0}^\infty  {\gamma _{ab}^{\left( m \right)}{\rho ^m}} , \quad \textrm{where} \quad \gamma _{ab}^{\left( m \right)} = \sum\limits_{n = 0}^m {{\partial _a}x_{\left( n \right)}^i{\partial _b}x_{\left( {m - n} \right)}^i} .
\label{eq:perturbation_gamma_expansion_coefficients}
\end{equation}
Obviously $\gamma _{ab}^{\left( 0 \right)} = \mathcal{G}_{ab}$. We also assume an expansion for the determinant of the induced metric of the form
\begin{equation}
\sqrt{\det\gamma} = \frac{\sqrt{\det \mathcal{G}}}{\rho^{d-2}} \sum\limits_{m = 0}^\infty  {\gamma_{\left( m \right)}{\rho ^m}} .
\label{eq:perturbation_det_g}
\end{equation}

The equation \eqref{eq:flow_homogenous_Hd} implies that the function $c$ is regular at $\rho = 0$. Therefore, we also assume an expansion for $c$ of the form
\begin{equation}
c = \sum\limits_{m = 0}^\infty  {c_{\left( m \right)}{\rho ^m}} .
\label{eq:perturbation_c_expansion}
\end{equation}

We recall that we have selected a particular parametrization of the minimal surface, so that the vector $\partial_\rho x^i$ is perpendicular to the vectors $\partial_a x^i$, i.e. $\partial_\rho x^i \partial_a x^i = 0$. Substituting the expansion \eqref{eq:perturbation_x_expansion} into this relation yields
\begin{equation}
\sum\limits_{m = 0}^\infty  {\sum\limits_{n = 0}^m {\left( {n + 1} \right)x_{\left( {n + 1} \right)}^i{\partial _a}x_{\left( {m - n} \right)}^i{\rho ^m}} }  = 0,
\end{equation}
implying that
\begin{equation}
\sum\limits_{n = 0}^m {\left( {n + 1} \right)x_{\left( {n + 1} \right)}^i{\partial _a}x_{\left( {m - n} \right)}^i} = 0,
\label{eq:perturbation_orthogonality}
\end{equation}
for any $m$. In what follows, we will refer to the constraints \eqref{eq:perturbation_orthogonality} as ``orthogonality conditions''.

Finally, the equation \eqref{eq:embedding_c_specoal_parametrization}, allows the connection between the expansion of $c$ and the expansion of the embedding functions. This equation assumes the form
\begin{equation}
\frac{1}{{{c^2}}} = 1 + \sum\limits_{m = 0}^\infty  {\sum\limits_{n = 0}^m {\left( {m - n + 1} \right)\left( {n + 1} \right)x_{\left( {n + 1} \right)}^i x_{\left( {m - n + 1} \right)}^i{\rho ^m}} } .
\label{eq:perturbation_invc2_expansion}
\end{equation}

We may proceed to solve perturbatively the equation \eqref{eq:flow_homogenous_Hd}. The expansions for $c$ and $\gamma$ are provided by equations \eqref{eq:perturbation_gamma_expansion_coefficients} and \eqref{eq:perturbation_invc2_expansion}. The parametrization freedom that could prohibit a unique solution to the equation is removed through the specific parametrization selection \eqref{eq:parametrization}, which is perturbatively expressed as \eqref{eq:perturbation_orthogonality}. Thus, it is a matter of algebra to solve the problem order by order.

\subsection{The Perturbative Solution}
\label{subsec:solution}
\subsubsection*{Order $\mathcal{O} \left( \rho^0 \right)$}
At leading order, the induced metric reads
\begin{equation}
\gamma_{ab} = \frac{\gamma_{ab}^{(0)}}{\rho^2} + \mathcal{O} \left(\frac{1}{\rho} \right) = \frac{\mathcal{G}_{ab}}{\rho^2} + \mathcal{O} \left(\frac{1}{\rho} \right) ,
\end{equation}
which means that
\begin{equation}
\sqrt{\det\gamma} = \frac{\sqrt{\det\mathcal{G}}}{\rho^{d-2}} + \mathcal{O} \left( \frac{1}{\rho^{d-3}} \right).
\end{equation}
Substituting this to the flow equation \eqref{eq:flow_homogenous_Hd} yields
\begin{equation}
\rho\partial_\rho\left(\frac{c_{\left( 0 \right)}}{\rho^{d-1}}\right) + \mathcal{O} \left( \frac{1}{\rho^{d - 2}} \right) = - \frac{d - 1}{c_{\left( 0 \right)} \rho^{d-1}} + \mathcal{O} \left( \frac{1}{\rho^{d - 2}} \right),
\end{equation}
which obviously implies that $c_{\left( 0 \right)} = 1$. Reading the equation \eqref{eq:perturbation_invc2_expansion} at leading order yields
\begin{equation}
c_{\left( 0 \right)} = 1 + x_{(1)}^i x_{(1)}^i,
\end{equation}
which implies that
\begin{equation}
x^i_{\left( 1 \right)} = 0 .
\end{equation}

\subsubsection*{Order $\mathcal{O} \left( \rho^1 \right)$}

The next order is rather trivial due to the fact that $x^i_{\left( 1 \right)} = 0$. The orthogonality condition \eqref{eq:perturbation_orthogonality} at leading order yields
\begin{equation}
x_{(1)}^i {\partial_a x_{(0)}^i} = 0,
\label{eq:perturbation_x1}
\end{equation}
which is trivially satisfied.

The equation \eqref{eq:perturbation_gamma_expansion_coefficients} at this order reads
\begin{equation}
\gamma _{ab}^{\left( 1 \right)} = {{\partial _a}x_{\left( 0 \right)}^i{\partial _b}x_{\left( 1 \right)}^i} +{{\partial _a}x_{\left( 1 \right)}^i{\partial _b}x_{\left( 0 \right)}^i} =0.
\end{equation}
Similarly, equation \eqref{eq:perturbation_invc2_expansion} implies that
\begin{equation}
c_{(1)} = - 2 x_{(2)}^i x_{(1)}^i = 0 ,
\end{equation}
and, thus, the flow equation \eqref{eq:flow_homogenous_Hd} is trivially satisfied to this order.

\subsubsection*{Order $\mathcal{O} \left( \rho^2 \right)$}

At next order, we receive new information from the orthogonality condition \eqref{eq:perturbation_orthogonality}, which reads,
\begin{equation}
x_{(2)}^i {\partial_a x_{(0)}^i} = 0,
\label{eq:orthogonality_O_rho}
\end{equation}
stating that the vector $x_{(2)}$ is perpendicular to the entangling surface, and thus, parallel to the normal vector $\mathcal{N}$.

At order $\mathcal{O} (\rho^2)$, the induced metric \eqref{eq:perturbation_gamma_expansion_coefficients} reads
\begin{equation}
\gamma _{ab}^{\left( 2 \right)} = {{\partial _a}x_{\left( 0 \right)}^i{\partial _b}x_{\left( 2 \right)}^i} +{{\partial _a}x_{\left( 2 \right)}^i{\partial _b}x_{\left( 0 \right)}^i} ,
\label{eq:perturbation_metric_O_2}
\end{equation}
due to the fact that $x_{\left( 1 \right)}^i = 0$. This implies that the determinant of the induced metric is given by
\begin{equation}
\sqrt{\det\gamma} = \frac{\sqrt{\det \mathcal{G}}}{\rho^{d-2}} \left( 1 + \gamma_{\left( 2 \right)} \rho^2 + \mathcal{O} \left( \rho^3 \right) \right) , \quad \text{where} \quad \gamma_{\left( 2 \right)} = \frac{1}{2}{\mathcal{G}}^{ab}\gamma_{ab}^{(2)} .
\label{eq:perturbation_det_O_rho^2}
\end{equation}
Using the expansion of the induced metric \eqref{eq:perturbation_metric_O_2}, together with \eqref{eq:orthogonality_O_rho} yields 
\begin{equation}
\gamma_{\left( 2 \right)} = - \mathcal{G}^{ab} x_{(2)}^i {\partial_a \partial_b x_{(0)}^i} .
\end{equation}

The expansion \eqref{eq:perturbation_invc2_expansion} at this order yields
\begin{equation}
c_{\left( 2 \right)} = - 2 x_{(2)}^i x_{(2)}^i .
\label{eq:perturbation_c_2}
\end{equation}
Plugging the expressions \eqref{eq:perturbation_det_O_rho^2} and \eqref{eq:perturbation_c_2} into the flow equation \eqref{eq:flow_homogenous_Hd} yields the relation
\begin{equation}
\gamma_{\left( 2 \right)} = \left( d - 2 \right) c_{\left( 2 \right)},
\label{eq:perturbation_flow_gamma_2}
\end{equation} 
and, thus, 
\begin{equation}
2 \left( d - 2 \right) x_{(2)}^i x_{(2)}^i = \mathcal{G}^{ab} x_{(2)}^i {\partial_a \partial_b x_{(0)}^i} .
\label{eq:perturbation_flow_2}
\end{equation}
Notice that this equation is satisfied for any $x_{(2)}$ when $d = 2$. In this case, the right hand side of the above equation vanishes, due to the fact that the entangling surface is zero-dimensional.

As we have already stated, the vector $x_{\left( 2 \right)}$ is parallel to the normal vector $\mathcal{N}$, i.e. $x_{\left( 2 \right)}^i = \sqrt{x_{\left( 2 \right)}^i x_{\left( 2 \right)}^i} \mathcal{N}$. Substituting this to \eqref{eq:perturbation_k_entangling} and using the orthogonality relation \eqref{eq:orthogonality_O_rho} yields
\begin{equation}
\mathcal{K}_{ab} = \frac{x_{(2)}^i {\partial_a \partial_b x_{(0)}^i}}{\sqrt{x_{(2)}^j x_{(2)}^j}} .
\label{eq:perturbation_second_fundamental_form}
\end{equation}
The mean curvature $\mathcal{K}$ equals
\begin{equation}
\mathcal{K} = \mathcal{G}^{ab} \mathcal{K}_{ab} = \frac{\mathcal{G}^{ab} x_{(2)}^i {\partial_a \partial_b x_{(0)}^i}}{\sqrt{x_{(2)}^j x_{(2)}^j}} = 2 \left( d - 2 \right) \sqrt{x_{(2)}^i x_{(2)}^i} ,
\label{eq:perturbation_mean curvature}
\end{equation}
due to the flow equation \eqref{eq:perturbation_flow_2}. It follows directly from equations \eqref{eq:perturbation_c_2} and \eqref{eq:perturbation_flow_gamma_2} that whenever $d > 2$,
\begin{equation}
x_{\left( 2 \right)}^i = - \frac{\mathcal{K}}{2 \left( d - 2 \right)} \mathcal{N}^i
\end{equation}
and
\begin{equation}
c_{\left( 2 \right)} = - \frac{\mathcal{K}^2}{2 \left( d - 2 \right)^2} , \quad \gamma_{\left( 2 \right)} = - \frac{\mathcal{K}^2}{2 \left( d - 2 \right)} .
\label{eq:perturbation_c_and_g_2}
\end{equation} 

\subsubsection*{Order $\mathcal{O} \left( \rho^3 \right)$}

The orthogonality condition at this order yields
\begin{equation}
x_{(3)}^i {\partial_a x_{(0)}^i} = 0.
\label{eq:orthogonality_O_rho^2}
\end{equation}

At order $\mathcal{O} \left( \rho^3 \right)$, the induced metric \eqref{eq:perturbation_gamma_expansion_coefficients} reads
\begin{equation}
\gamma _{ab}^{\left( 3 \right)} = {{\partial _a}x_{\left( 0 \right)}^i{\partial _b}x_{\left( 3 \right)}^i} +{{\partial _a}x_{\left( 3 \right)}^i{\partial _b}x_{\left( 0 \right)}^i} ,
\label{eq:perturbation_metric_O_3}
\end{equation}
due to the fact that $x_{\left( 1 \right)}^i = 0$.
The determinant of the induced metric is given by
\begin{equation}
\sqrt{\det\gamma} = \frac{\sqrt{\det \mathcal{G}}}{\rho^{d-2}} \left( 1 + \gamma_{\left( 2 \right)} \rho^2 + \gamma_{\left( 3 \right)} \rho^3 + \mathcal{O} \left( \rho^4 \right) \right) , \quad \text{with} \quad \gamma_{\left( 3 \right)} = \frac{1}{2}{\mathcal{G}}^{ab}\gamma_{ab}^{(3)} .
\label{eq:perturbation_det_O_rho^3}
\end{equation}
The relation \eqref{eq:orthogonality_O_rho^2}, implies that the vector $x_{\left( 3 \right)}$ is perpendicular to the entangling surface. We recall that the same holds for $x_{\left( 2 \right)}$ due to \eqref{eq:orthogonality_O_rho}. Therefore both $x_{\left( 2 \right)}$ and $x_{\left( 3 \right)}$ are parallel to the normal vector $\mathcal{N}$, and, thus, to each other, i.e., $x_{\left( 3 \right)}^i = \sqrt {{{x_{\left( 3 \right)}^j x_{\left( 3 \right)}^j}}} x_{\left( 2 \right)}^i / \sqrt {{{x_{\left( 2 \right)}^k x_{\left( 2 \right)}^k}}}$. This equation combined with \eqref{eq:orthogonality_O_rho^2}, \eqref{eq:perturbation_metric_O_3} and \eqref{eq:perturbation_flow_2} implies that
\begin{equation}
\gamma_{\left( 3 \right)} = - \sqrt {\frac{{x_{\left( 3 \right)}^j x_{\left( 3 \right)}^j}}{{x_{\left( 2 \right)}^k x_{\left( 2 \right)}^k}}} \mathcal{G}^{ab} x_{(2)}^i {\partial_a \partial_b x_{(0)}} = - 2 \left(d - 2 \right) \sqrt {{{x_{\left( 2 \right)}^i x_{\left( 2 \right)}^i}}{{x_{\left( 3 \right)}^j x_{\left( 3 \right)}^j}}} .
\label{eq:perturbation_detg_3}
\end{equation}

Furthermore, the equation \eqref{eq:perturbation_c_2} implies that
\begin{equation}
c_{\left( 3 \right)} = - 6 x_{(2)}^i x_{(3)}^i = - 6 \sqrt {{{x_{\left( 2 \right)}^i x_{\left( 2 \right)}^i}}{{x_{\left( 3 \right)}^j x_{\left( 3 \right)}^j}}} .
\label{eq:perturbation_c_3}
\end{equation}

To this order the flow equation \eqref{eq:flow_homogenous_Hd} yields
\begin{equation}
\left( 2 d - 5 \right) c_{\left( 3 \right)} = 3 \gamma_{\left( 3 \right)} \quad \text{or} \quad \left( d - 3 \right) \sqrt{{x_{\left( 3 \right)}^i x_{\left( 3 \right)}^i}} = 0 .
\label{eq:perturbation_flow_3}
\end{equation}
This means that the flow equation is satisfied automatically to this order if $d = 3$ for any $x_{\left( 3 \right)}$ parallel to $\mathcal{N}$. On the contrary for any $d \geq 4$ the above equation implies that.
\begin{equation}
x_{\left( 3 \right)}^i = 0,
\end{equation}
which further implies that $c_{\left( 3 \right)} = 0$ and $\gamma_{\left( 3 \right)} = 0$.

\subsubsection*{Order $\mathcal{O} \left( \rho^4 \right)$}

The orthogonality relation \eqref{eq:perturbation_orthogonality} at this order reads
\begin{equation}
2 x_{(4)}^i {\partial_a x_{(0)}^i} + x_{(2)}^i {\partial_a x_{(2)}^i} = 0 .
\label{eq:orthogonality_O_rho_3}
\end{equation}

The induced metric \eqref{eq:perturbation_gamma_expansion_coefficients} reads
\begin{equation}
\gamma _{ab}^{\left( 4 \right)} = {{\partial _a}x_{\left( 0 \right)}^i {\partial _b}x_{\left( 4 \right)}^i} + {{\partial _a}x_{\left( 2 \right)}^i{\partial _b}x_{\left( 2 \right)}^i} + {{\partial _a}x_{\left( 4 \right)}^i{\partial _b}x_{\left( 0 \right)}^i} ,
\label{eq:perturbation_metric_O_4}
\end{equation}
due to the fact that $x_{\left( 1 \right)}^i = 0$. The determinant of the induced metric is given by
\begin{equation}
\sqrt{\det\gamma} = \frac{\sqrt{\det \mathcal{G}}}{\rho^{d-2}} \left( 1 + \gamma_{\left( 2 \right)} \rho^2 + \gamma_{\left( 3 \right)} \rho^3 + \gamma_{\left( 4 \right)} \rho^4 + \mathcal{O} \left( \rho^5 \right) \right) ,
\label{eq:perturbation_det_O_rho^4}
\end{equation}
where
\begin{multline}
2 \gamma_{\left( 4 \right)} = \gamma_{\left( 2 \right)}^2 - 2 \mathcal{G}^{ab} \mathcal{G}^{cd}{{\partial _a}x_{\left( 0 \right)}^i {\partial _c}x_{\left( 2 \right)}^i} {{\partial _b}x_{\left( 0 \right)}^j {\partial _d}x_{\left( 2 \right)}^j} \\
+ \mathcal{G}^{ab} \left( {{\partial _a}x_{\left( 2 \right)}^k {\partial _b}x_{\left( 2 \right)}^k} + 2 {{\partial _a}x_{\left( 0 \right)}^l {\partial _b}x_{\left( 4 \right)}^l} \right) .
\label{eq:det_O_rho^4}
\end{multline}
Using the equations \eqref{eq:perturbation_second_fundamental_form} and \eqref{eq:perturbation_mean curvature}, the second term in \eqref{eq:det_O_rho^4} assumes the form
\begin{equation}
\mathcal{G}^{ab} \mathcal{G}^{cd}{{\partial _a}x_{\left( 0 \right)}^i {\partial _c}x_{\left( 2 \right)}^i} {{\partial _b}x_{\left( 0 \right)}^j {\partial _d}x_{\left( 2 \right)}^j} = \frac{\mathcal{K}^2 \mathcal{K}_{ab}\mathcal{K}^{ab}}{4 \left( d - 2 \right)^2}.
\end{equation}
Using equations \eqref{eq:orthogonality_O_rho_3} and \eqref{eq:perturbation_mean curvature}, the third term in \eqref{eq:det_O_rho^4} assumes the form
\begin{equation}
\begin{split}
\mathcal{G}^{ab} &\left( {{\partial _a}x_{\left( 2 \right)}^k {\partial _b}x_{\left( 2 \right)}^k} + 2 {{\partial _a}x_{\left( 0 \right)}^l {\partial _b}x_{\left( 4 \right)}^l} \right) \\
&= - \mathcal{G}^{ab} \left( x_{(2)}^k {\partial_a \partial_b x_{(2)}^k} + 2 x_{(4)}^l {\partial_a \partial_b x_{(0)}^l} \right) \\
&= - \mathcal{G}^{ab} \left( {\frac{1}{2} {\partial_a \partial_b \left( x_{(2)}^i x_{(2)}^i \right) } - {\partial_a x_{(2)}^k} {\partial_b x_{(2)}^k} + 2 x_{(4)}^l {\partial_a \partial_b x_{(0)}^l}} \right) \\
&= - \mathcal{G}^{ab} \left( \frac{{\partial_a \partial_b} \mathcal{K}^2}{8 \left( d - 2 \right)^2} - {\partial_a x_{(2)}^k} {\partial_b x_{(2)}^k} + 2 x_{(4)}^l {\partial_a \partial_b x_{(0)}^l} \right).
\end{split}
\label{eq:C_3_term}
\end{equation}

The vector $x_{(2)}$ is parallel to the normal vector. Thus, the vectors $\left\{ x_{(2)} , {\partial_a x_{(0)}} \right\}$ form a basis. We decompose the vectors $x_{(4)}$ and ${\partial_a x_{(2)}}$ into this basis,
\begin{align}
{\partial_a x_{(2)}^i} &= A_a x_{(2)}^i + A_a^c {\partial_c x_{(0)}^i} ,
\label{eq:decomposition_partial_x_2} \\
x_{(4)}^i &= f x_{(2)}^i + f^c {\partial_c x_{(0)}^i} .
\label{eq:decomposition_x_4}
\end{align}
Taking the inner product of \eqref{eq:decomposition_partial_x_2} with $x_{(2)}$ and utilizing \eqref{eq:orthogonality_O_rho}, together with \eqref{eq:perturbation_mean curvature} leads to $A_a = \frac{\partial_a \mathcal{K}}{\mathcal{K}} $. Similarly multiplying \eqref{eq:decomposition_partial_x_2} with ${\partial_b x_{(0)}^i}$ and utilizing \eqref{eq:orthogonality_O_rho}, \eqref{eq:perturbation_second_fundamental_form} and \eqref{eq:perturbation_mean curvature} yields $A_a^c = - \frac{\mathcal{G}^{ce} \mathcal{K} \mathcal{K}_{ae}}{2 \left( d - 2 \right)} $, and, thus,
\begin{equation}
{\partial_a x_{(2)}^i} = \frac{\partial_a \mathcal{K}}{\mathcal{K}} x_{(2)}^i - \frac{\mathcal{G}^{ce} \mathcal{K} \mathcal{K}_{ae}}{2 \left( d - 2 \right)} {\partial_c x_{(0)}^i} .
\label{eq:decomposition_partial_x_2_b}
\end{equation}
In the same spirit, we plug the decomposition \eqref{eq:decomposition_x_4} into the orthogonality relation \eqref{eq:orthogonality_O_rho_3} and after some algebra we arrive at $f^c = - \frac{\mathcal{G}^{cb} \mathcal{K} {\partial_b \mathcal{K}}}{8 \left( d - 2 \right)^2} $, and, thus, 
\begin{equation}
x_{(4)}^i = f x_{(2)}^i - \frac{\mathcal{G}^{cb} \mathcal{K} {\partial_b \mathcal{K}}}{8 \left( d - 2 \right)^2} {\partial_c x_{(0)}^i} .
\label{eq:decomposition_x_4_b}
\end{equation}

Now we can compute the quantities that appear in \eqref{eq:C_3_term}. The equation \eqref{eq:decomposition_x_4_b} implies that
\begin{equation}
\mathcal{G}^{ab} x_{(4)}^i {\partial_a \partial_b x_{(0)}^i} = \frac{\mathcal{K}^2 f}{2 \left( d - 2 \right)} - \frac{\mathcal{G}^{ab} \mathcal{G}^{ce} \mathcal{K} {\partial_e \mathcal{K}}}{8 \left( d - 2 \right)^2} {\partial_c x_{(0)}^i} {\partial_a \partial_b x_{(0)}^i} .
\end{equation}
It can be easily shown that ${\partial_c x_{(0)}^i} {\partial_a \partial_b x_{(0)}^i} = \mathcal{G}_{cd} \mathit{\Gamma}^{d}_{ab} $, where $\mathit{\Gamma}^{d}_{ab}$ are the Christoffel symbols with respect to the induced metric $\mathcal{G}$ of the entangling surface, namely, $\mathit{\Gamma}^{d}_{ab} = \frac{1}{2} \mathcal{G}^{de} \left( {\partial_a \mathcal{G}_{be}} + {\partial_b \mathcal{G}_{ae}} - {\partial_e \mathcal{G}_{ab}} \right) $. Thus,
\begin{equation}
\mathcal{G}^{ab} x_{(4)}^i {\partial_a \partial_b x_{(0)}^i} = \frac{f \mathcal{K}^2}{ 2 \left( d - 2 \right)} - \frac{\mathcal{K} \mathcal{G}^{ab} \mathit{\Gamma}^{d}_{ab} {\partial_d \mathcal{K}}}{8 \left( d - 2 \right)^2} .
\end{equation}
Similarly, the equation \eqref{eq:decomposition_partial_x_2_b} implies
\begin{equation}
{\partial_a x_{(2)}^i} {\partial_b x_{(2)}^i} = \frac{{\partial_a \mathcal{K}} {\partial_b \mathcal{K}}}{4 \left( d - 2 \right)^2} + \frac{\mathcal{K}^2 \mathcal{G}^{ cd} \mathcal{K}_{ad} \mathcal{K}_{bc}}{4 \left( d - 2 \right)^2} .
\end{equation}
Putting everything together, the third term in \eqref{eq:det_O_rho^4} is written as
\begin{equation}
\mathcal{G}^{ab} \left( {{\partial _a}x_{\left( 2 \right)}^k {\partial _b}x_{\left( 2 \right)}^k} + 2 {{\partial _a}x_{\left( 0 \right)}^l {\partial _b}x_{\left( 4 \right)}^l} \right) = - \frac{f \mathcal{K}^2}{d-2} + \frac{\mathcal{K}^2 \mathcal{K}_{ab} \mathcal{K}^{ab}} {4 \left( d - 2 \right)^2} - \frac{\mathcal{K} \Box \mathcal{K}}{4 \left( d - 2 \right)^2} ,
\end{equation}
where $\Box = \mathcal{G}^{ab} \nabla_a \nabla_b$, while the covariant derivatives are taken with respect to the induced metric of the entangling surface. Summing up, the equation \eqref{eq:det_O_rho^4} assumes the form
\begin{equation}
\gamma_{\left( 4 \right)} = \frac{\mathcal{K}^4}{8 \left( d - 2 \right)^2} - \frac{f \mathcal{K}^2}{2 \left( d - 2 \right)} - \frac{\mathcal{K}^2 \mathcal{K}_{ab} \mathcal{K}^{ab}}{8 \left( d - 2 \right)^2} - \frac{\mathcal{K} \Box \mathcal{K}}{8 \left( d - 2 \right)^2}.
\end{equation}

The equation \eqref{eq:perturbation_invc2_expansion} implies that
\begin{equation}
c_{ \left( 4 \right) } = 6 \left( x_{(2)}^i x_{(2)}^i \right)^2 - 8 x_{(4)}^i x_{(2)}^i - \frac{9 x_{(3)}^i x_{(3)}^i}{2}.
\end{equation}
Using \eqref{eq:perturbation_mean curvature}, together with \eqref{eq:decomposition_x_4_b} leads to
\begin{equation}
c_{ \left( 4 \right) } = \frac{3 \mathcal{K}^4}{8 \left( d - 2 \right)^4} - \frac{2 f \mathcal{K}^2}{\left( d - 2 \right)^2} - \frac{9 x_{(3)}^i x_{(3)}^i}{2}.
\label{eq:c_4}
\end{equation}

Expanding the flow equation \eqref{eq:flow_homogenous_Hd} to this order yields
\begin{equation}
\left( d - 5 \right) \left( c_{ \left( 4 \right) } + c_{ \left( 2 \right) } \gamma_{ \left( 2 \right) } + \gamma_{ \left( 4 \right) } \right) = \left( d - 1 \right) \left(\gamma_{ \left( 4 \right) } - c_{ \left( 2 \right) } \gamma_{ \left( 2 \right) } + (c_{ \left( 2 \right) })^2 - c_{ \left( 4 \right) } \right),
\end{equation}
or
\begin{equation}
\frac{\mathcal{K}^4}{2 \left( d - 2 \right)^4} + \frac{\left( d - 4 \right) f \mathcal{K}^2}{\left( d - 2 \right)^2} - \frac{\mathcal{K}^2 \mathcal{K}_{ab} \mathcal{K}^{ab}}{4 \left( d - 2 \right)^2} - \frac{\mathcal{K} \Box \mathcal{K}}{4 \left( d - 2 \right)^2} + \frac{9 \left( d - 3 \right) x_{(3)}^i x_{(3)}^i}{2} = 0.
\end{equation}
We recall that $x_{(3)}^i = 0$ for any $d \neq 3$. It follows that the last term is always vanishing, allowing the re-expression of the last equation as
\begin{equation}
4 \left( d - 4 \right) f = - \frac{ 2 \mathcal{K}^2}{\left( d - 2 \right)^2} + {\mathcal{K}_{ab} \mathcal{K}^{ab}} + \frac{\Box \mathcal{K}}{\mathcal{K}} ,
\label{eq:perturbation_flow_4}
\end{equation}
This implies that in any number of dimensions except for the case $d = 4$, the quantity $f$, and, thus $x_{(4)}$ is completely determined by the local characteristics of the part of the entangling surface that we are expanding around. When, $d \neq 4$, the above equation directly determines $f$ and it implies that
\begin{equation}
x_{(4)}^i = \frac{\mathcal{K}}{8 \left( d - 2 \right) \left( d - 4 \right)} \left( - \frac{ 2 \mathcal{K}^2}{\left( d - 2 \right)^2} + {\mathcal{K}_{ab} \mathcal{K}^{ab}} + \frac{\Box \mathcal{K}}{\mathcal{K}} \right) \mathcal{N}^i - \frac{\mathcal{G}^{cb} \mathcal{K} {\partial_b \mathcal{K}}}{8 \left( d - 2 \right)^2} {\partial_c \mathcal{X}^i} ,
\end{equation}
\begin{equation}
\gamma_{ \left( 4 \right) } = \frac{\left( d - 3 \right) \mathcal{K}^2}{ 4 \left( d - 2 \right)^2 \left( d - 4 \right)} \left( \frac{ \left( d - 3 \right)^2 + 1}{2 \left( d - 2 \right) \left( d - 3 \right)}\mathcal{K}^2 - \mathcal{K}_{ab} \mathcal{K}^{ab} - \frac{\Box \mathcal{K}}{\mathcal{K}} \right).
\label{eq:perturbation_g_4}
\end{equation}
and
\begin{equation}
c_{ \left( 4 \right) } = \begin{cases}
\frac{\mathcal{K}^2}{2 \left( d - 2 \right)^2 \left( d - 4 \right)} \left( \frac{3 d - 4}{4 \left( d - 2 \right)^2} \mathcal{K}^2 - \mathcal{K}_{ab} \mathcal{K}^{ab} - \frac{\Box \mathcal{K}}{\mathcal{K}} \right) , & d \geq 5 ,\\
- \frac{\mathcal{K}^4}{8}+\frac{\mathcal{K}\Box \mathcal{K}}{2} - \frac{9 x_{(3)}^i x_{(3)}^i}{2}, & d = 3.
\end{cases}
\label{eq:perturbation_c_4}
\end{equation}
When $d = 4$, we have shown that the component of $x_{\left( 4 \right)}$ that is perpendicular to the entangling  surface is undetermined. In this case, the flow equation \eqref{eq:perturbation_flow_4} reduces to
\begin{equation}
- \frac{\mathcal{K}^2}{2} + {\mathcal{K}_{ab} \mathcal{K}^{ab}} + \frac{\Box \mathcal{K}}{\mathcal{K}} = 0 ,
\label{eq:solution_d4_constraint}
\end{equation}
which is a constraint for the entangling surface. When the entangling surface does not satisfy this constraint, there are implications for the form of the expansion of the embedding functions. We will return to this issue in section \ref{subsec:solution_Neumann}.

\subsection{The Neumann Boundary Condition in the Perturbative Expansion}
\label{subsec:solution_Neumann}

At all orders higher than the first one, we found that at order $d$ the equations cannot completely determine the solution. This is due to the fact that at this order the Neumann boundary condition enters into the solution. Let us first analyse this behaviour at the orders that have already been studied in section \ref{subsec:solution}, using some clarifying examples, before we proceed to make some more general comments.

\subsubsection*{$d = 2$}
When $d = 2$, i.e. in the case of AdS$_3$, we found that the flow equation \eqref{eq:perturbation_flow_2} is satisfied for any $x_{(2)}$ parallel to $\mathcal{N}$. At this number of dimensions, it is easy to show that this behaviour is due to the fact that the Neumann boundary condition for the differential equation \eqref{eq:flow_homogenous_Hd}, which is determined by the existence of other disconnected boundaries, enters into the solution at the second order. In pure AdS$_3$, all static minimal surfaces are either semicircles of the form $\left( x - x_0 \right)^2 = R^2 - \rho^2$, or semi-infinite straight lines $x = x_0$, if there is no other boundary. Expanding the semi-circle solution around one of the two boundary points, e.g. $x = x_0 + R \equiv x_1$, yields
\begin{equation}
x = x_1 - \frac{1}{2R} \rho^2 + \mathcal{O} \left( \rho^3 \right).
\end{equation}
Thus, indeed, the second order term depends on the parameter $R$, i.e. on the existence of a part of the entangling surface (in this case entangling points), which is disconnected from the part of the entangling surface around which we expand our solution (in this case $x = x_1$). Notice also that this term vanishes at the limit $R \to \infty$, i.e in the case that there is no other disconnected segment of the entangling surface.

\subsubsection*{$d = 3$}
\begin{figure}[ht]
\vspace{10pt}
\begin{center}
\begin{picture}(100,37)
\put(0,0){\includegraphics[width = 0.5\textwidth]{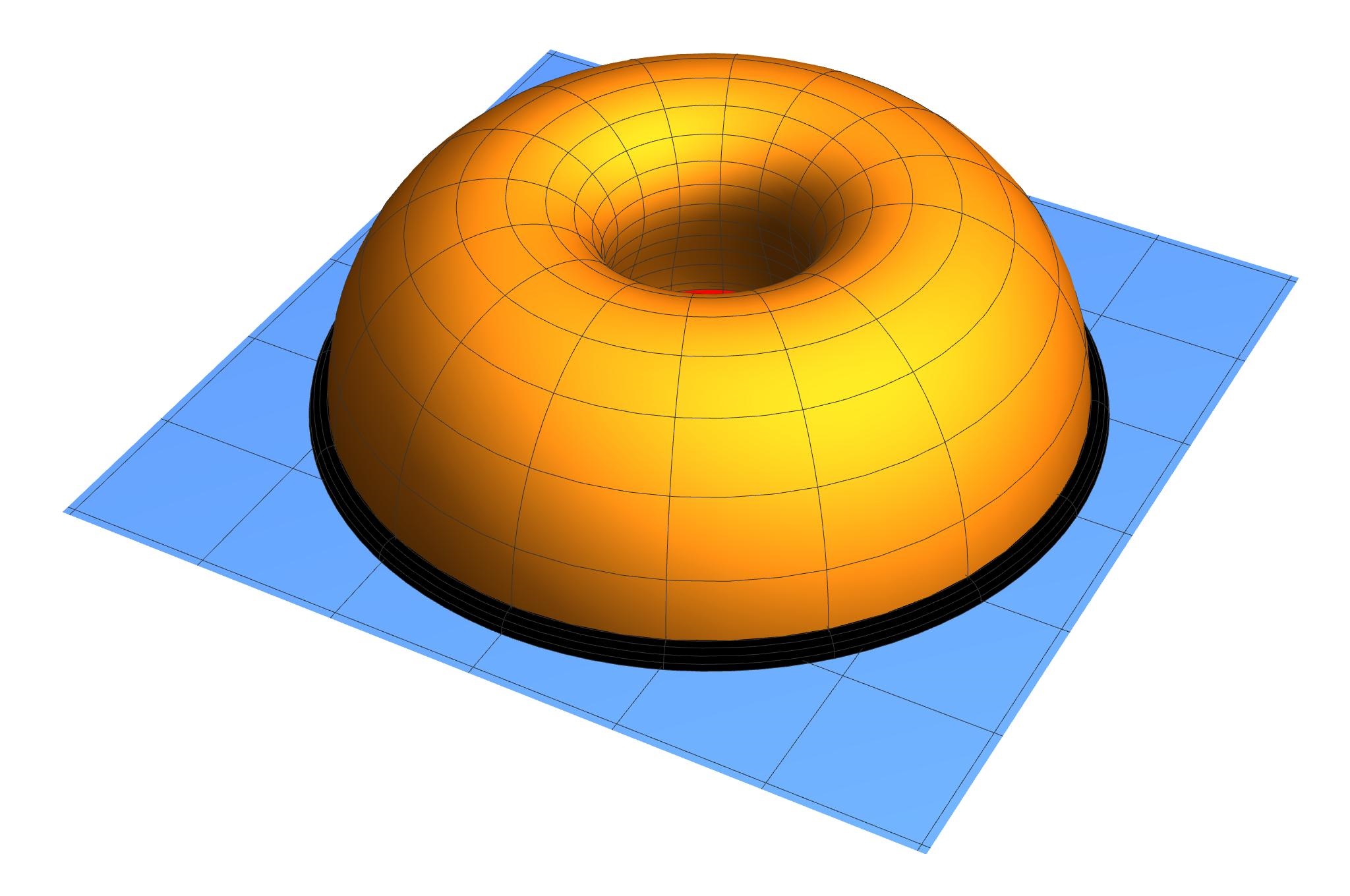}}
\put(50,0){\includegraphics[width = 0.5\textwidth]{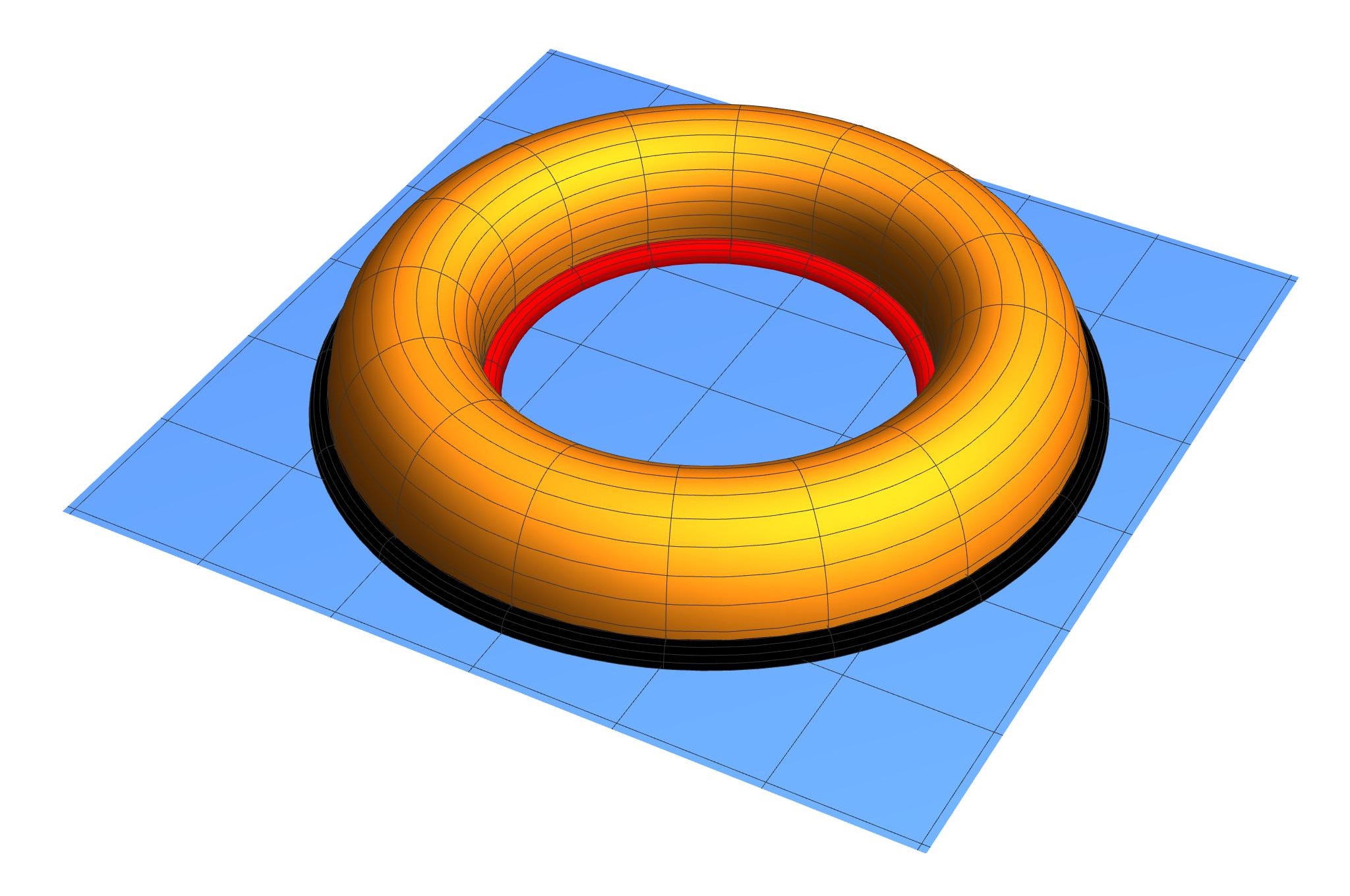}}
\end{picture}
\end{center}
\vspace{-10pt}
\caption{Two catenoids whose corresponding entangling curves do not coincide but they share a common part, which is plotted as the black curve.}
\vspace{5pt}
\label{fig:catenoids}
\end{figure}

When $d = 3$, we found that the flow equation \eqref{eq:perturbation_flow_3} is satisfied for any vector $x_{(3)}$ parallel to $\mathcal{N}$. This property is similar to what occurred at the previous order for $d = 2$. Again, at this order, the Neumann boundary condition enters into the solution. A nice clarifying example for this behaviour is the case of catenoid minimal surfaces in H$^3$, since they correspond to a disconnected entangling surface, which comprises of two concentric circles. These surfaces are parametrized by \cite{Pastras:2016vqu}
\begin{equation}
\rho  = \sqrt {\frac{{3{e_2}}}{{\wp \left( u \right) + 2{e_2}}}} {e^{ - {\varphi _1}\left( {u;a_1} \right)}} , \quad \left| \vec x \right| = \sqrt {\frac{{\wp \left( u \right) - {e_2}}}{{\wp \left( u \right) + 2{e_2}}}} {e^{ - {\varphi _1}\left( {u;a_1} \right)}} ,
\label{eq:catenoid}
\end{equation}
where
\begin{equation}
{\varphi _1}\left( {u;a{\kern 1pt} } \right) = \frac{1}{2}\ln \left( { - \frac{{\sigma \left( {u + {a_1}} \right)}}{{\sigma \left( {u - {a_1}} \right)}}} \right) - \zeta \left( {{a_1}} \right)u .
\end{equation}
The functions $\wp$, $\zeta$ and $\sigma$ are the Weierstrass elliptic function and the related quasi-periodic functions, respectively, with moduli ${g_2} = \frac{{{E^2}}}{3} + 1$ and ${g_3} =  - \frac{E}{3}\left( {\frac{{{E^2}}}{9} + \frac{1}{2}} \right)$. The quantity $e_2$ is the intermediate root of the related cubic polynomial, namely $e_2 = \frac{E}{6}$. The parameter $a_1$ assumes a specific value so that $\wp \left( a_1 \right) = - 2 e_2$ and finally the parameter $E$ may assume any positive value. The catenoid is covered for a full real period $2 \omega_1$ of the Weierstrass elliptic function. Considering the segment $u \in \left[0 , 2\omega_1 \right]$ or $u \in \left[ - 2\omega_1 , 0 \right]$, the catenoid is anchored at the boundary at two concentric circles, one with radius $R$ and another one, whose radius equals $R \, {\exp \left[{ \mp {\mathop{\rm Re}\nolimits} \left( {\zeta \left( {{\omega _1}} \right){\alpha _1} - \zeta \left( {{a_1}} \right){\omega _1}} \right)} \right] }$, hence it depends on the value of the parameter $E$. Figure \ref{fig:catenoids} shows two catenoid minimal surfaces whose entangling curves do not coincide. However they comprise of two concentric circles, one of whom is common.

Expanding the catenoid solution around the part of the entangling surface, which is the circle of radius $R$, is equivalent to expanding the embedding functions around $u = 0$. This yields
\begin{equation}
\begin{split}
\rho  &=  \pm R\sqrt {\frac{E}{2}} \left( {u - \frac{E}{6}{u^3} + \mathcal{O}\left( {{u^4}} \right)} \right) , \\
\left| \vec x \right| &= R\left( {1 - \frac{E}{4}{u^2} + \frac{1}{6}\sqrt {\frac{E}{2}} {u^3} + \mathcal{O}\left( {{u^4}} \right)} \right) ,
\end{split}
\end{equation}
implying
\begin{equation}
\left| \vec x \right| = R - \frac{1}{{2R}}{\rho^2} \pm \frac{1}{{3E{R^2}}}{\rho^3} + \mathcal{O}\left( {{\rho^4}} \right) .
\end{equation}
It is evident that the coefficient of the $\rho^2$ term depends solely on the geometry of the part of the entangling curve around which we are expanding, i.e. on the radius $R$. Actually it has exactly the right value as described by the formula \eqref{eq:perturbation_mean curvature}, namely $\left| x_{\left( 2 \right)} \right| = \frac{1}{{2R}} = \frac{\mathcal{K}}{2 \left( d - 2 \right)}$. On the other hand, the coefficient of the $\rho^3$ depends on the parameter $E$, i.e. on the position of the other circle that constitutes the entangling surface. Notice again that at the limit where the other circle disappears, i.e. $E \to \infty$, this term vanishes. Although they look quite different, the two catenoids plotted in figure \ref{fig:catenoids} have the same expansion up to order $\rho^2$.

The catenoids do not exhaust the freedom of the selection of the Neumann boundary condition. They are just the solutions that preserve the rotational symmetry, which at this expansion is equivalent to the selection of a $x_{(3)}$ with constant magnitude. Keeping the same Dirichlet boundary conditions and selecting a more general Neumann boundary condition would lead to a minimal surface corresponding to a disconnected entangling curve comprised of a circle and another curve, which would not be a circle.

\subsubsection*{$d = 4$}
When $d = 4$, we have shown that the component of $x_{\left( 4 \right)}$ that is perpendicular to the entangling  surface is undetermined. This is the expected freedom due to the potential existence of other disconnected parts of the entangling surface. However, in this case, the flow equation reduces to \eqref{eq:solution_d4_constraint},
which is a constraint for the entangling surface. This constraint may hold (e.g. in the case of a spherical entangling surface where the two principal curvatures are $\kappa_1 = \kappa_2 = 1 / R$) in which case, the expansion we have performed is valid. On the contrary, the expansion \eqref{eq:perturbation_x_expansion} is inconsistent when this constraint does not hold (e.g. in the case of a cylindrical entangling surface where the two principal curvatures are $\kappa_1 = 1 / R$, $\kappa_2 = 0$). In the following we will show that in such a case this problem is resolved via the introduction of a $\rho^4 \ln \rho$ term in the expansion of the embedding functions, which does not alter the perturbation theory at lower orders. As expected, the component of $x_{\left( 4 \right)}$ that is perpendicular to the entangling surface remains undetermined by the flow equation, and, thus, it is determined by the Neumann boundary condition.

\subsubsection*{Arbitrary Number of Dimensions}
Let us investigate the general structure of the flow equation \eqref{eq:flow_homogenous_Hd} in the perturbation theory that we developed, in order to understand how the equation determines the embedding functions of the minimal surface order by order. Using the notation \eqref{eq:perturbation_det_g} and \eqref{eq:perturbation_c_expansion} and introducing a similar notation for $1 / c$, the flow equation at order $n$ reads
\begin{equation}
\left( {n - d + 1} \right)\sum\limits_{k = 0}^n {{c_{\left( k \right)}}{\gamma _{\left( {n - k} \right)}}}  + \left( {d - 1} \right)\sum\limits_{k = 0}^n {{{\left( {\frac{1}{c}} \right)}_{\left( k \right)}}{\gamma _{\left( {n - k} \right)}}}  = 0 .
\label{eq:perturbation_flow_n_order}
\end{equation}

First, we need to understand what is the highest order term of the embedding functions that appears in $c_{\left( n \right)}$ and $\gamma_{\left( n \right)}$. Trivially, equation \eqref{eq:perturbation_gamma_expansion_coefficients} implies that in $\gamma_{\left( n \right)}$, this is $x_{\left( n \right)}^i$. The equation \eqref{eq:perturbation_invc2_expansion} naively suggests that the highest order term that appears in $c_{\left( n \right)}$ is $x_{\left( n + 1 \right)}^i$; however this is multiplied with $x_{\left( 1 \right)}^i$, which vanishes. Therefore, the highest order term that appears in $c_{\left( n \right)}$ is also $x_{\left( n \right)}^i$. It follows that naturally, the $n$-th order of the perturbation theory determines the $x_{\left( n \right)}^i$ term of the embedding functions.

The equation \eqref{eq:perturbation_invc2_expansion} implies that
\begin{equation}
{\left( {\frac{1}{{{c^2}}}} \right)_{\left( n \right)}} = \sum\limits_{k = 0}^n {\left( {n - k + 1} \right)\left( {k + 1} \right)x_{\left( {k + 1} \right)}^ix_{\left( {n - k + 1} \right)}^i}  = 4nx_{\left( 2 \right)}^ix_{\left( n \right)}^i + \mathcal{F}\left( {x_{\left( {m < n} \right)}^i} \right) ,
\end{equation}
where $\mathcal{F}\left( {x_{\left( {m < n} \right)}^i} \right)$ denotes a function of the terms, which are of order lower than $n$. We use this notation without implying that $\mathcal{F}$ is some specific function, but in the same fashion that we use the symbol $\mathcal{O} \left( \rho^n \right)$ to denote the terms of order $\rho^n$ and higher in an expansion. The above equation implies that
\begin{equation}
{\left( {\frac{1}{c}} \right)_{\left( n \right)}} = 2nx_{\left( 2 \right)}^i x_{\left( n \right)}^i + \mathcal{F}\left( {x_{\left( {m < n} \right)}^i} \right),\quad {c_{\left( n \right)}} =  - 2nx_{\left( 2 \right)}^ix_{\left( n \right)}^i + \mathcal{F}\left( {x_{\left( {m < n} \right)}^i} \right) .
\label{eq:perturbation_order_n_c_n}
\end{equation}
In a similar manner
\begin{equation}
\gamma _{ab}^{\left( n \right)} = {\partial _a}x_{\left( n \right)}^i{\partial _b}x_{\left( 0 \right)}^i + {\partial _a}x_{\left( 0 \right)}^i{\partial _b}x_{\left( n \right)}^i + \mathcal{F} _{ab} \left( {x_{\left( {m < n} \right)}^i} \right) ,
\end{equation}
which implies that
\begin{equation}
{\gamma _{\left( n \right)}} = \frac{1}{2}{\mathcal{G}^{ab}}\gamma _{ab}^{\left( n \right)} + \mathcal{F}\left( {x_{\left( {m < n} \right)}^i} \right) = {\mathcal{G}^{ab}}{\partial _a}x_{\left( n \right)}^i{\partial _b}x_{\left( 0 \right)}^i + \mathcal{F} \left( {x_{\left( {m < n} \right)}^i} \right) .
\label{eq:perturbation_order_n_gamma_n}
\end{equation}

The orthogonality condition \eqref{eq:perturbation_orthogonality} implies that
\begin{equation}
x_{\left( n \right)}^i{\partial _a}x_{\left( 0 \right)}^i = \mathcal{F}_a \left( {x_{\left( {m < n} \right)}^i} \right),
\label{eq:perturbation_order_n_orthogonality}
\end{equation}
which allows the re-expression of \eqref{eq:perturbation_order_n_gamma_n} as
\begin{equation}
{\gamma _{\left( n \right)}} =  - {\mathcal{G}^{ab}}x_{\left( n \right)}^i{\partial _a}{\partial _b}x_{\left( 0 \right)}^i + \mathcal{F}\left( {x_{\left( {m < n} \right)}^i} \right) .
\end{equation}

We use the fact that the vector $x_{\left( 2 \right)}$ is perpendicular to the entangling surface, and thus, the vectors $\left\{ x_{\left( 2 \right)} , \partial_a x_{\left( 0 \right)} \right\}$ form a base. We decompose $x_{\left( n \right)}$ in this base as
\begin{equation}
x_{\left( n \right)}^i = {X_{\left( n \right)}}x_{\left( 2 \right)}^i + X_{\left( n \right)}^a{\partial _a}x_{\left( 0 \right)}^i .
\label{eq:perturbation_order_n_decomposition}
\end{equation}
Notice that actually, only the perpendicular component ${X_{\left( n \right)}}$ is a new degree of freedom that appears at this order. All other components are completely determined by the solution at lower orders through the orthogonality condition \eqref{eq:perturbation_order_n_orthogonality}. Indeed, substituting \eqref{eq:perturbation_order_n_decomposition} in \eqref{eq:perturbation_order_n_orthogonality} yields $X_{\left( n \right)}^a{\partial _a}x_{\left( 0 \right)}^i{\partial _b}x_{\left( 0 \right)}^i = X_{\left( n \right)}^a{\mathcal{G}_{ab}} = {\mathcal{F}_b}\left( {x_{\left( {m < n} \right)}^i} \right)$, which directly implies that $X_{\left( n \right)}^a = {\mathcal{G}^{ab}}\mathcal{F}_b \left( {x_{\left( {m < n} \right)}^i} \right) = \mathcal{F}^a \left( {x_{\left( {m < n} \right)}^i} \right)$.

Substituting \eqref{eq:perturbation_order_n_decomposition} in \eqref{eq:perturbation_order_n_c_n} and \eqref{eq:perturbation_order_n_gamma_n} and taking advantage of the equation \eqref{eq:perturbation_flow_2} yields
\begin{align}
{\gamma _{\left( n \right)}} &=  - 2\left( {d - 2} \right){X_{\left( n \right)}}x_{\left( 2 \right)}^ix_{\left( 2 \right)}^i - X_{\left( n \right)}^c{G^{ab}} {\partial _a}{\partial _b}x_{\left( 0 \right)}^i {\partial _c}x_{\left( 0 \right)}^i + \mathcal{F}\left( {x_{\left( {m < n} \right)}^i} \right) , \label{eq:perturbation_order_n_gamma_n_b}\\
{\left( {\frac{1}{c}} \right)_{\left( n \right)}} &= 2n{X_{\left( n \right)}}x_{\left( 2 \right)}^ix_{\left( 2 \right)}^i + \mathcal{F}\left( {x_{\left( {m < n} \right)}^i} \right),\quad {c_{\left( n \right)}} =  - 2n{X_{\left( n \right)}}x_{\left( 2 \right)}^ix_{\left( 2 \right)}^i + \mathcal{F}\left( {x_{\left( {m < n} \right)}^i} \right) . \label{eq:perturbation_order_n_c_n_b}
\end{align}

We isolate the terms $k=0$ and $k=n$ of the equation \eqref{eq:perturbation_flow_n_order}, which are the only ones that contain $x_{\left( n \right)}^i$, bearing in mind that $c_{\left( 0 \right)} = 1$ and $\gamma_{\left( 0 \right)} = 1$. Then, this equation assumes the form
\begin{multline}
\left( {n - d + 1} \right){c_{\left( n \right)}} + \left( {d - 1} \right){\left( {\frac{1}{c}} \right)_{\left( n \right)}} + n{\gamma _{\left( n \right)}} \\
=  - \left( {n - d + 1} \right)\sum\limits_{k = 1}^{n - 1} {{c_{\left( k \right)}}{\gamma _{\left( {n - k} \right)}}}  - \left( {d - 1} \right)\sum\limits_{k = 1}^{n - 1} {{{\left( {\frac{1}{c}} \right)}_{\left( k \right)}}{\gamma _{\left( {n - k} \right)}}}  = \mathcal{F}\left( {x_{\left( {m < n} \right)}^i} \right) .
\end{multline}
Finally, substituting \eqref{eq:perturbation_order_n_gamma_n_b} and \eqref{eq:perturbation_order_n_c_n_b} in the above equation yields
\begin{equation}
2\left( {d - n} \right){X_{\left( n \right)}}x_{\left( 2 \right)}^ix_{\left( 2 \right)}^i - X_{\left( n \right)}^c{G^{ab}} {\partial _a}{\partial _b}x_{\left( 0 \right)}^i {\partial _c}x_{\left( 0 \right)}^i = \mathcal{F}\left( {x_{\left( {m < n} \right)}^i} \right) .
\end{equation}
This clearly implies that at order $d$, the flow equation does not determine the component of $x_{\left( n \right)}^i$ that is perpendicular to the entangling surface. This component is determined by the Neumann boundary condition. As we already commented above, the components of $x_{\left( n \right)}$ that are parallel to the entangling surface, i.e. the coefficients $X_{\left( n \right)}^c$, are completely determined by the lower order terms of the solution through the orthogonality condition. Therefore, at $n = d$, the solution reduces to a constraint for the solution at lower orders than $d$. We have already seen this as the equation \eqref{eq:solution_d4_constraint} in the case $d = 4$.

\begin{figure}[ht]
\vspace{10pt}
\begin{center}
\begin{picture}(45,42)
\put(0,0){\includegraphics[width = 0.5\textwidth]{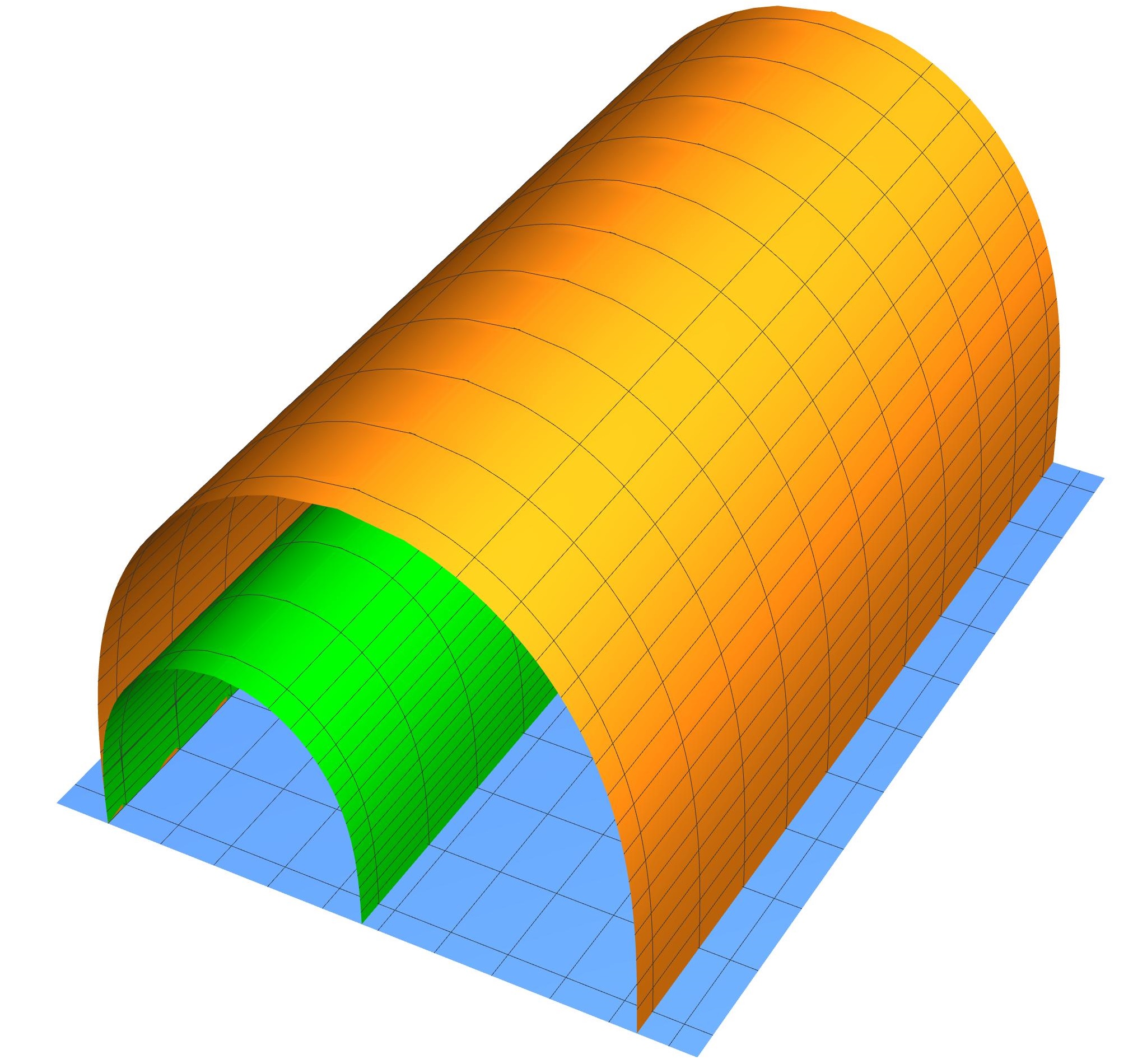}}
\end{picture}
\end{center}
\vspace{-10pt}
\caption{Two minimal surfaces corresponding to strip regions. The entangling curves do not coincide but they share a common part.}
\vspace{5pt}
\label{fig:strips}
\end{figure}

An indicative example of this behaviour is the minimal surface that corresponds to a strip region. It is well-known that this minimal surface satisfies the equation
\begin{equation}
\frac{{dx \left( \rho \right)}}{{d\rho}} =  \frac{\rho^{d - 1}}{{\sqrt {{R^{2\left(d - 1\right)}} - {\rho^{2\left(d - 1\right)}}} }} ,
\end{equation}
where $R$ is the maximum value of the holographic coordinate on the minimal surface, which is related to the width of the strip region. It follows that the expansion of this minimal surface reads
\begin{equation}
x = x_1 + \frac{\rho^d}{d R^{d-1}} + \mathcal{O} \left( \rho^{d+1} \right) .
\end{equation}
This means that all strip minimal surfaces that share one edge of the strip region, such as those plotted in figure \ref{fig:strips}, have an identical expansion up to order $\mathcal{O} \left( \rho^{d-1} \right)$. Of course in this special case, all these terms vanish, as a consequence of the fact that the curvature of the entangling surface vanishes.

The above imply that all terms of the solution of odd order smaller than $d$ vanish. Actually, this holds for all odd orders, whenever $d$ is even. We can show this iteratively. Assuming that $n$ is odd and all odd orders up to $n$ are vanishing, then, all the functions $\mathcal{F}\left( {x_{\left( {m < n} \right)}^i} \right)$ that appeared in the above derivation are actually vanishing, since they constitute of a sum of products of odd and even lower ordered terms. Since we have already showed that the first order vanishes, all terms of odd order vanish when $d$ is even, whereas when $d$ is odd, all terms of odd order smaller than $d$ are vanishing. Furthermore, the consistency condition that emerges at order $d$, where $d$ is odd is trivially satisfied, as both the left hand side and the right hand side are vanishing.

\subsubsection*{Logarithmic Terms in the Expansion of the Embedding Functions}
\label{subsec:solution_log}

We have seen that at order $d$, the flow equation cannot determine the component of $x_{d}$ that is perpendicular to the entangling surface, which is determined by the Neumann boundary condition, but it rather reduces to a constraint for the terms of the solution of order smaller than $d$. These terms have already been determined by the perturbation theory at lower orders and can be expressed in terms of the extrinsic geometry of the entangling surface. Thus, at order $d$ the flow equation reduces to a constraint for the geometry of the entangling surface. When $d$ is odd, this constraint is trivially satisfied, as a consequence of the fact that all lower order odd terms vanish. In this case or when $d$ is even and the entangling surface satisfies the constraint, no consistency problem occurs in our expansion. The question that remains to be answered is what happens when $d$ is even and the constraint is not satisfied. In such a case, the regular Taylor expansion of the embedding functions that we used is incomplete and one has to include logarithmic terms at orders $d$ and higher.

Let us introduce a logarithmic term at order $d$. Then, the expansion of the embedding functions of the minimal surface will read
\begin{equation}
{x^i}\left( {\rho ;{u^a}} \right) = \sum\limits_{m = 0}^d {x_{\left( m \right)}^i\left( {{u^a}} \right){\rho ^m}}  + \tilde x_{\left( d \right)}^i\left( {{u^a}} \right){\rho ^d}\ln \rho  + \mathcal{O}\left( {{\rho ^{d + 1}}} \right) .
\label{eq:logarithm_xi}
\end{equation}

The orthogonality condition \eqref{eq:perturbation_orthogonality} up to this order reads
\begin{multline}
\sum\limits_{m = 0}^{d - 1} {\sum\limits_{n = 0}^m {\left( {n + 1} \right)x_{\left( {n + 1} \right)}^i{\partial _a}x_{\left( {m - n} \right)}^i{\rho ^m}} }  + d \tilde x_{\left( d \right)}^i{\partial _a}x_{\left( 0 \right)}^i{\rho ^{d - 1}} \\
+ \tilde x_{\left( d \right)}^i{\partial _a}x_{\left( 0 \right)}^i{\rho ^{d - 1}}\ln \rho  + \mathcal{O}\left( {{\rho ^d}} \right) = 0 .
\label{eq:logarithm_orthogonality}
\end{multline}
This clearly implies that
\begin{equation}
\tilde x_{\left( d \right)}^i{\partial _a}x_{\left( 0 \right)}^i = 0 ,
\label{eq:logarithmi_xdtilde_orth}
\end{equation}
meaning that the vector $\tilde x_{\left( d \right)}$ is perpendicular to the entangling surface, i.e.
\begin{equation}
\tilde x_{\left( d \right)}^i = {{\tilde X}_{\left( d \right)}}x_{\left( 2 \right)}^i .
\end{equation}
The equation \eqref{eq:logarithmi_xdtilde_orth} implies that the second term of the equation \eqref{eq:logarithm_orthogonality} vanishes. Thus, the rest of the orthogonality conditions remain unaltered by the introduction of the logarithmic term.

Using the expansion \eqref{eq:logarithm_xi} to find the expansion of the determinant of the induced metric yields
\begin{equation}
\sqrt {\det \gamma }  = \frac{{\sqrt {\det G} }}{{{\rho ^{d - 2}}}}\left( {\sum\limits_{m = 0}^d {{\gamma _{\left( m \right)}}{\rho ^m}}  + {{\tilde \gamma }_{\left( d \right)}}{\rho ^d}\ln \rho  + \mathcal{O}\left( {{\rho ^{d + 1}}} \right)} \right) ,
\end{equation}
where
\begin{equation}
{{\tilde \gamma }_{\left( d \right)}} = {\mathcal{G}^{ab}}{\partial _a}x_{\left( 0 \right)}^i{\partial _b}\tilde x_{\left( d \right)}^i 
\label{eq:logarithm_gamma_tilde}
\end{equation}
and ${\gamma }_{\left( d \right)}$ are given by the same expressions as in the expansion without the logarithmic term.

Substituting the expansion \eqref{eq:logarithm_xi} into \eqref{eq:perturbation_invc2_expansion}, we find that $1 / c^2$ has an expansion of the form
\begin{equation}
\frac{1}{{{c^2}}} = \sum\limits_{m = 0}^d {{{\left( {\frac{1}{{{c^2}}}} \right)}_{\left( m \right)}}{\rho ^m}}  + \left( {\frac{1}{{{c^2}}}} \right)_{\left( d \right)}^\prime{\rho ^d}\ln \rho + \mathcal{O}\left( {{\rho ^{d + 1}}} \right) ,
\end{equation}
where
\begin{equation}
{\left( {\frac{1}{{{c^2}}}} \right)_{\left( d \right)}} = 4x_{\left( 2 \right)}^i\left( {dx_{\left( d \right)}^i + \tilde x_{\left( d \right)}^i} \right) + \mathcal{F}\left( {x_{\left( {m < d} \right)}^i} \right),\quad \left( {\frac{1}{{{c^2}}}} \right)_{\left( d \right)}^\prime = 4dx_{\left( 2 \right)}^i\tilde x_{\left( d \right)}^i
\end{equation}
and all other coefficients ${{\left( {\frac{1}{{{c^2}}}} \right)}_{\left( m \right)}}$, with $m < d$ remain unaltered by the introduction of the logarithmic term. Adopting a similar notation for the expansions of $c$ and $1/c$, the above implies that
\begin{align}
{\left( {\frac{1}{c}} \right)_{\left( d \right)}} = 2x_{\left( 2 \right)}^i\left( {dx_{\left( d \right)}^i + \tilde x_{\left( d \right)}^i} \right) + \mathcal{F}\left( {x_{\left( {m < d} \right)}^i} \right),&\quad \left( {\frac{1}{c}} \right)_{\left( d \right)}^\prime = 2dx_{\left( 2 \right)}^i\tilde x_{\left( d \right)}^i , \label{eq:logarithm_invc_prime}\\
{c_{\left( d \right)}} =  - 2x_{\left( 2 \right)}^i\left( {dx_{\left( d \right)}^i + \tilde x_{\left( d \right)}^i} \right) + \mathcal{F}\left( {x_{\left( {m < d} \right)}^i} \right),&\quad {c_{\left( d \right)}^\prime} =  - 2dx_{\left( 2 \right)}^i\tilde x_{\left( d \right)}^i . \label{eq:logarithm_c_prime}
\end{align}

We may now substitute the expansions of the determinant of the induced metric and $c$ into the flow equation \eqref{eq:flow_homogenous_Hd}. All equations at orders smaller than $d$ remain unaltered, whereas at order $d$ we will get two equations: one from the coefficient of $\rho^d$ and one from the coefficient of $\rho^d \ln \rho$. The latter reads
\begin{equation}
{{\tilde \gamma }_{\left( d \right)}} + {{\tilde c}_{\left( d \right)}} + \left( {d - 1} \right)\left( {{{\tilde \gamma }_{\left( d \right)}} + \left( {\frac{1}{c}} \right)_{\left( d \right)}^\prime} \right) = 0 .
\end{equation}
Using equations \eqref{eq:logarithm_gamma_tilde}, \eqref{eq:logarithm_invc_prime} and \eqref{eq:logarithm_c_prime}, the above equation assumes the form
\begin{equation}
{G^{ab}}{\partial _a}x_{\left( 0 \right)}^i{\partial _b}\tilde x_{\left( d \right)}^i =  - 2\left( {d - 2} \right)x_{\left( 2 \right)}^i\tilde x_{\left( d \right)}^i .
\end{equation}
This equation is always true as a result of \eqref{eq:logarithmi_xdtilde_orth} and \eqref{eq:perturbation_flow_2}.

The equation obtained from the coefficient of $\rho^d$ is
\begin{equation}
{c_{\left( d \right)}} + \left( {d - 1} \right){\left( {\frac{1}{c}} \right)_{\left( d \right)}} + d{\gamma _{\left( d \right)}} = \mathcal{F}\left( {x_{\left( {m < d} \right)}^i} \right) .
\end{equation}
Implementing \eqref{eq:logarithm_gamma_tilde}, \eqref{eq:logarithm_invc_prime} and \eqref{eq:logarithm_c_prime}, the above equation assumes the form
\begin{equation}
2\left( {d - 2} \right){{\tilde X}_{\left( d \right)}}x_{\left( 2 \right)}^ix_{\left( 2 \right)}^i = \mathcal{F}\left( {x_{\left( {m < d} \right)}^i} \right) .
\end{equation}
As before introducing the logarithmic term, the component ${X}_{\left( d \right)}$ does not appear and remains undetermined by the flow equation. This component is determined by the Neumann boundary condition. However, this equation ceases being a constraint for the lower order terms, but it determines the component ${\tilde X}_{\left( d \right)}$. For example at $d = 4$, we get
\begin{equation}
{\tilde X}_{\left( 4 \right)} = \frac{ \mathcal{K}^2}{8} - \frac{\mathcal{K}_{ab} \mathcal{K}^{ab}}{4} - \frac{\Box \mathcal{K}}{4 \mathcal{K}} .
\end{equation}

The introduction of the logarithmic term solved the consistency problem. Without that term, we had one free parameter and one equation that did not contain this free parameter and could be inconsistent. After the introduction of the logarithmic term, we have two free parameters and two equations. One of the parameters still does not appear in the equations, but one of the latter is always satisfied, no matter what the value of the other parameter is.

In a straightforward manner, at orders higher than $d$, one has to include logarithmic terms. As the order increases higher powers of logarithms may be necessary. The equations though are going to be always as many as the free parameters, allowing the perturbative determination of the embedding functions at arbitrary order.

\setcounter{equation}{0}
\section{The Divergent Terms of Entanglement Entropy in Pure AdS Space}
\label{sec:entropy}

When Einstein gravity is considered in the bulk, the entanglement entropy is given by the original Ryu-Takayanagi formula, i.e.
\begin{equation}
S_{\mathrm{EE}} = \frac{A}{8 \pi G} ,
\end{equation}
where $A$ is the area of the minimal surface in the bulk, which is anchored at the entangling surface.

We cutoff the minimal surface at $\rho = 1 / \Lambda$. Then, in the specific parametrization \eqref{eq:parametrization} that we have used, the area of the minimal surface is given by the expression
\begin{equation}
A\left( \Lambda \right) = \int_{1 / \Lambda}^{\rho_{\max}} \, d\rho \int \, d^{d - 2}u \sqrt{\det\Gamma} = \int_{1 / \Lambda}^{\rho_{\max}} \, d\rho \int \, d^{d - 2}u \frac{\sqrt{f(\rho) \det\gamma}}{c} ,
\label{eq:area:general}
\end{equation}
where $\rho_{\max}$ is the maximum value of the holographic coordinate on the minimal surface. When we consider minimal surfaces that correspond to connected entangling surfaces, this $\rho_{\max}$ indeed assumes a given value\footnote{Even for connected surfaces it is possible that more than one local maxima of the holographic coordinate exist. In such a case, there are saddle points of the minimal surface. The topology of the intersection of the minimal surface with the constant-$r$ planes changes at the value of the holographic coordinate where a saddle point appears. At the level of the flow equation \eqref{eq:flow_Hd}, a saddle point is a point where the function $a \left( \rho ; u^a \right)$ becomes infinite and the normal vector $n$ is not well-defined. In such cases, the integral formula \eqref{eq:area:general} has to be split to patches separated by the saddle points, see also the discussion in section \ref{subsec:saddle}.}, e.g. in the case of a spherical entangling surface of radius $R$, $\rho_{\max} = R$. When we consider minimal surfaces that correspond to non-connected entangling surfaces, the situation is more complicated, since one has to run the flow from each disconnected part and arrange a smooth matching of the initially disconnected parts of the minimal surface. In any case, the details of $\rho_{\max}$ affect only the term which is constant in the cutoff expansion. Although this constant term is of great physical significance, here we focus on the divergent terms. It is evident that the expansion we developed in the previous section can be used to systematically derive these terms.

In pure AdS$_{d+1}$ in \Poincare coordinates, $f \left( \rho \right) = 1 / \rho^2$, thus the equation \eqref{eq:area:general} assumes the form
\begin{equation}
A \left( \Lambda \right) = \int_{1 / \Lambda}^{\rho_{\max}} d\rho \int d^{d-2}u \frac{\sqrt{\det\gamma}}{\rho c}.
\end{equation}
Using the flow equation \eqref{eq:flow_homogenous_Hd}, we obtain
\begin{equation}
A \left( \Lambda \right) = - \frac{1}{d - 1} \left[ \left. \int d^{d-2}u \left( c \sqrt{\det \gamma} \right) \right|_{\rho = 1 / \Lambda}^{\rho = \rho_{\max}} - \int_{1 / \Lambda}^{\rho_{\max}} d\rho \int \, d^{d - 2}u \frac{c\sqrt{\det\gamma}}{\rho} \right].
\label{eq:area}
\end{equation}

Finally, incorporating the expansions \eqref{eq:perturbation_det_g} and \eqref{eq:perturbation_c_expansion} the above equation assumes the form
\begin{multline}
A\left( \Lambda  \right) =  - \frac{1}{{d - 1}}\sum\limits_{n = 0}^\infty  {\left[ {\left( {\sum\limits_{m = 0}^n {\int {{d^{d - 2}}u\sqrt {\det G} {c_{\left( m \right)}}{\gamma _{\left( {n - m} \right)}}} } } \right)} \right.} \\
{\times \left. {\left( {\left. {\frac{1}{{{\rho ^{d - n - 2}}}}} \right|_{\rho  = 1 / \Lambda }^{\rho  = {\rho _{\max }}} - \int_{1 / \Lambda} ^{{\rho _{\max }}} {\frac{{d\rho }}{{{\rho ^{d - n - 1}}}}} } \right)} \right]} .
\end{multline}
This clarifies that the divergent terms are determined by the expansion of the minimal surface up to order $d - 2$. The Neumann boundary condition, i.e. the non-local properties of the entangling surface, affect the terms of order $d$ and higher. It follows that all divergent terms depend solely on the local characteristics of the entangling surface. Furthermore, we have shown that all terms of odd order lower than $d$ vanish. Therefore, when $d$ is odd,
\begin{multline}
A\left( \Lambda  \right) = \frac{1}{{d - 1}}\sum\limits_{n = 0}^{\left( {d - 3} \right)/2} {\left[ {\left( {\sum\limits_{m = 0}^{\left( {d - 3} \right)/2} {\int {{d^{d - 2}}u\sqrt {\det G} {c_{\left( {2m} \right)}}{\gamma _{\left( {2n - 2m} \right)}}} } } \right)\frac{{d - 2n - 1}}{{d - 2n - 2}}{\Lambda ^{d - 2n - 2}}} \right]} \\
\textrm{ + non-divergent terms} ,
\end{multline}
whereas, when $d$ is even
\begin{multline}
A\left( \Lambda  \right) = \frac{1}{{d - 1}}\sum\limits_{n = 0}^{\left( {d - 4} \right)/2} {\left[ {\left( {\sum\limits_{m = 0}^{\left( {d - 4} \right)/2} {\int {{d^{d - 2}}u\sqrt {\det G} {c_{\left( {2m} \right)}}{\gamma _{\left( {2n - 2m} \right)}}} } } \right)\frac{{d - 2n - 1}}{{d - 2n - 2}}{\Lambda ^{d - 2n - 2}}} \right]}  \\
+ \frac{1}{{d - 1}}\left( {\sum\limits_{m = 0}^{\left( {d - 2} \right)/2} {\int {{d^{d - 2}}u\sqrt {\det G} {c_{\left( {2m} \right)}}{\gamma _{\left( {d - 2m - 2} \right)}}} } } \right)\ln \Lambda \textrm{ + non-divergent terms} .
\end{multline}

We adopt the notation
\begin{equation}
A\left( \Lambda  \right) = {a_0}\ln \Lambda  + \sum\limits_{n = 1}^{d - 2} {{a_n}{\Lambda ^n}} \textrm{ + non-divergent terms} .
\label{eq:area_notation}
\end{equation}
The leading divergence is the usual ``area law'' term. For any $d \geq 3$, the relevant coefficient is
\begin{equation}
a_{d - 2} = \frac{1}{d - 2} \int d^{d-2}u \, \sqrt{\det \mathcal{G}} = \frac{1}{d - 2} \mathcal{A} ,
\end{equation}
where $\mathcal{A}$ is the area of the entangling surface.

For any $d \geq 4$, there is at least one more divergent term. Using \eqref{eq:perturbation_c_and_g_2}, we find that the coefficient of this term equals
\begin{equation}
a_{d - 4} = \begin{cases}
- \frac{d - 3}{2 \left( d - 2 \right)^2 \left( d - 4 \right)} \int d^{d-2} u \,\sqrt{\det \mathcal{G}} \mathcal{K}^2 , & d \geq 4 , \\
- \frac{1}{8} \int d^2 u \, \sqrt{\det \mathcal{G}} \mathcal{K}^2 , & d = 4 .
\end{cases}
\end{equation}
At $d = 4$, this term is the universal logarithmic term. The value of its coefficient is in agreement with \cite{Solodukhin:2008dh}.

The next diverging correction to the area appears whenever $d \geq 6$. Reading equations \eqref{eq:perturbation_c_and_g_2}, \eqref{eq:perturbation_g_4} and \eqref{eq:perturbation_c_4}, we find
\begin{equation}
c_{(4)} + c_{(2)} \gamma_{(2)} + \gamma _{(4)} = \frac{d - 1}{4 \left( d - 2 \right)^2 \left( d - 4 \right)} \left[ \frac{d^2 - 5 d + 8}{2 \left( d - 2 \right)^2} \mathcal{K}^4 - \mathcal{K}^2 \mathcal{K}_{ab} \mathcal{K}^{ab} - \mathcal{K} \Box \mathcal{K} \right] .
\end{equation}
Therefore
\begin{equation}
a_{d - 6} = \begin{cases}
\frac{d - 5}{4 \left( d - 2 \right)^2 \left( d - 4 \right) \left( d - 6 \right)} \int d^{d-2} u \,\sqrt{\det \mathcal{G}} \left[ \frac{d^2 - 5 d + 8}{2 \left( d - 2 \right)^2} \mathcal{K}^4 - \mathcal{K}^2 \mathcal{K}_{ab} \mathcal{K}^{ab} - \mathcal{K} \Box \mathcal{K} \right] , & d \geq 6 , \\
\frac{1}{128} \int d^4 u \, \sqrt{\det \mathcal{G}} \left[ \frac{7}{16} \mathcal{K}^4 - \mathcal{K}^2 \mathcal{K}_{ab} \mathcal{K}^{ab} - \mathcal{K} \Box \mathcal{K} \right] , & d = 6 .
\end{cases}
\end{equation}
At $d=6$ this is a universal logarithmic term. It is in agreement with the results of \cite{Safdi:2012sn}, where the logarithmic term is expressed in terms of both the intrinsic and extrinsic geometry of the entangling surface. Our result is expressed in terms of the extrinsic geometry of the entangling surface solely and it has a quite simple expression.

Let us verify the above in the simple case of a spherical entangling surface of radius $R$. In this case $\mathcal{K} = \frac{d - 2}{R}$, $\mathcal{K}_{ab} \mathcal{K}^{ab} = \frac{d-2}{R^2}$ and $\Box \mathcal{K} = 0$.
Thus,
\begin{equation}
a_{d - 4} = \begin{cases}
- \frac{\left( d - 3 \right) \mathcal{A}_{d - 2}} {2 \left( d - 4 \right) R^2 } , & d \geq 4 , \\
- \frac{\mathcal{A}_{2}}{2 R^2} , & d = 4 ,
\end{cases} \quad
a_{d - 6} = \begin{cases}
\frac{\left( d - 3 \right) \left( d - 5 \right) \mathcal{A}_{d - 2}} {8 \left( d - 6 \right) R^4 } , & d \geq 6 , \\
\frac{3 \mathcal{A}_{4}}{8 R^4} , & d = 6 ,
\end{cases}
\label{eq:terms_ad}
\end{equation}
where $\mathcal{A}_{d}$ is the area of a $d$-dimensional sphere of radius $R$. The minimal surface, which corresponds to a spherical entangling surface, is analytically known, hence the above coefficients can be calculated directly. This task is performed in appendix \ref{sec:terms_simple}. The result of the direct calculation, which is provided by equations \eqref{eq:sphere_an}, \eqref{eq:sphere_a0} and \eqref{eq:sphere_a_d} is in perfect agreement with the perturbatively calculated coefficients above.

\setcounter{equation}{0}
\section{Discussion}
\label{sec:discussion}

Since the initial formulation of the Ryu-Takayanagi conjecture \cite{Ryu:2006bv,Ryu:2006ef}, which connects the entanglement entropy in the boundary theory to the area of minimal surfaces in the bulk, the study of minimal surfaces in asymptotically AdS spaces has received a great interest. The problem of the specification of a minimal surface in AdS for given boundary data presents great difficulty due to the non-linearity of the equations which are obeyed by the minimal surfaces. Actually, very few minimal surfaces are explicitly known; most of the related literature focuses on those that correspond to spherical entangling surfaces or strip regions on the boundary. An example of non-trivial minimal surfaces with explicit expressions is the family of the elliptic minimal surfaces in AdS$_4$ \cite{Pastras:2016vqu}, which includes the helicoids, the catenoids and the cusps. More general minimal surfaces are known in a more abstract, less handy form in terms of hyperelliptic functions \cite{Ishizeki:2011bf,Kruczenski:2013bsa}.

In this work, instead of relying on exact minimal surfaces, which necessarily correspond to specific entangling surfaces, we follow a different approach. First, we describe the minimal surface as a geometric flow of the entangling surface towards the interior of the bulk. In this language, the evolving entangling surface traces the minimal surface, in the same sense that a string traces its world-sheet. Then, we solve this flow equation perturbatively around the boundary, obtaining an expression for the minimal surface that corresponds to any smooth entangling surface.

The solution to the flow equation presents a specific dependence on the boundary conditions. Since it is a second order equation with respect to the holographic coordinate, two boundary conditions are required in order to uniquely specify a solution. The Dirichlet boundary condition is obviously the form of the entangling surface at the boundary. The second one is a Neumann-type boundary condition. Similarly to all second order differential equations, the Neumann boundary condition can also be expressed as a second Dirichlet boundary condition; it depends on the existence of other disconnected parts of the entangling surface, i.e. on non-local characteristics of the latter. Assuming that the bulk is AdS$_{d + 1}$, the solution does not depend on the Neumann boundary condition at any order smaller than $d$. All smaller orders are completely determined by the Dirichlet condition, i.e. the local characteristics of the entangling surface.

It turns out that the terms of order lower than $d$ in this perturbative solution of the flow equation are those which determine \emph{all} the divergent terms of the holographic entanglement entropy, including the universal logarithmic term in odd bulk spacetime dimensions. Thus, all the divergent terms depend only on the local characteristics of the entangling surface, such as its curvature. The perturbative solution to the flow equation constitutes a systematic method for the determination of these terms.

In this work, we found the three most divergent terms in pure AdS$_{d + 1}$ spaces, solely in terms of the extrinsic geometry of the entangling surface. These include simple expressions for the universal logarithmic terms both in AdS$_5$ and AdS$_7$, which are in agreement with the literature \cite{Solodukhin:2008dh,Safdi:2012sn}. Therein, these terms are calculated through the use of an ansatz dictated by the conformal symmetry. The purely geometric method, which we have applied here, verifies these results, without any assumptions. Moreover, it simplifies the obtained expressions and extends them to the polynomially divergent terms.

Our method has a number of obvious direct uses and generalizations. It is well known that minimal surfaces, which correspond to entangling curves with non-smooth points, such as conical or wedge singularities \cite{Klebanov:2012yf,Myers:2012vs} or more complicated logarithmic spiral ones \cite{Pastras:2017fsy}, generate new terms in the expansion of the holographic entanglement entropy that do not emerge for smooth entangling surfaces. These new terms include universal terms, which are proportional to the logarithm of the UV cutoff in even bulk dimensions and to its square in odd bulk dimensions. The coefficients of these terms can be related to the central charges of the dual CFT \cite{Myers:2012vs,Bueno:2015xda,Bueno:2015rda}. The machinery of the geometric flow which describes the minimal surface can be directly applied to the case of singular entangling surfaces in order to provide simple analytic expressions for all these terms in an arbitrary number of dimensions. In this language, the singular points are simply singularities in the Dirichlet boundary data (e.g. conical and wedge singularities are delta function singularities of the extrinsic curvature of the entangling surface) and therefore such terms can be studied in a unified fashion with the terms that emerge in the case of smooth entangling surfaces.

Whenever the CFT has an Einstein gravity holographic dual, the central charges are proportional to each other at leading order in the rank of the gauge group of the boundary theory. In effect, their contributions to the universal term are not discernible. For general higher derivative gravitational duals, the central charges cease being proportional to each other. These setups are very interesting, since they allow the study of a broader class of CFTs with unequal central charges. Since the central charges can be distinguished, one can in principle obtain a formula for the coefficient of the universal logarithmic term that is valid for arbitrary values of the central charges, independently of the specific gravitational dual. 

In view of this, the generalization of the Ryu–Takayanagi prescription for the calculation of the holographic entanglement entropy for more general gravitational theories is required. The correspondence between the entanglement entropy and the entropy of topological black holes \cite{Casini:2011kv}, motivates the use of Wald’s functional instead of the area, for this purpose. Yet, this naive guess does not give the right answer \cite{Hung:2011xb}. There are plenty of works in the literature that discuss the functional that should be minimized. This discussion was initiated in the context of Lovelock gravity in \cite{Hung:2011xb} and \cite{deBoer:2011wk}. The simplest case of Lovelock gravity, namely Gauss-Bonnet gravity, is discussed extensively in \cite{Bhattacharyya:2013jma}, whereas general curvature square theories are studied in \cite{Camps:2013zua}. Even more general theories whose Lagrangians depend on contractions of the Riemann tensor were treated in \cite{Dong:2013qoa}. Yet, the picture is far from clear since these results were debated \cite{Bhattacharyya:2014yga}, while various subtleties are not well understood \cite{Miao:2014nxa,Astaneh:2014sma,Astaneh:2014wxg,Astaneh:2015tea}.

In the present work, we have worked out a purely geometric approach to this problem, which is generalizable for any functional, via the appropriate modification of equation \eqref{eq:flow}. In effect, our approach enables a holographic calculation, which does not rely on any ansatz for the expected result.

Our geometric flow method can also be easily adapted to the study other bulk geometries, which have very interesting applications, via the appropriate adaptation of equations \eqref{eq:flow_Hd} or \eqref{eq:flow_homogenous_Hd}. A first trivial example would be the study of the AdS black hole geometry, which would allow the specification of thermal corrections to the holographic entanglement entropy. However, the form of the AdS Black hole metric
\begin{equation}
{ds}^2 = - \left( k^2 r^2 + 1 - \frac{C}{r^{d - 2}} \right) {dt}^2 + \left( k^2 r^2 + 1 - \frac{C}{r^{d - 2}} \right)^{-1} {dr}^2 + r^2 {d \Omega}^2
\end{equation}
implies that deviations from the pure AdS case appear at order $d$ in the perturbation theory, hence they do not affect the divergent terms of the holographic entanglement entropy. This is not surprising since the thermal contributions are not expected to be relevant in the UV of the theory. The same holds for any perturbation of the pure AdS geometry, which obeys Dirichlet boundary conditions. This becomes obvious via the Fefferman-Graham expansion of such geometries. Among these geometries, one of particular interest is the AdS soliton background, which is related to confinement-deconfinement phase transitions in the boundary. Indeed, it is known that it is the constant \emph{non-divergent} term of entanglement entropy that plays the role of a quantum order parameter \cite{Nishioka:2006gr,Klebanov:2007ws}.

On the other hand, one may study the geometry generated by probe branes, which corresponds to massive deformations of the boundary field theory. These geometries do not possess AdS asymptotics and are known to generate new universal logarithmic terms, associated with the mass scale introduced in the boundary theory \cite{Hung:2011ta,Chang:2013mca,Karch:2014ufa,Jones:2015twa,Taylor:2017zzo}.

Furthermore, it would be particularly interesting to study systems with Fermi surfaces, as in such systems, the leading divergence of the entanglement entropy is not the usual ``area law'' term, but it is enhanced from $\Lambda^{d - 2}$ to $\Lambda^{d - 2} \ln \Lambda$ \cite{Ogawa:2011bz,Shaghoulian:2011aa}. Our method is appropriate for the specification of all divergent terms and additionally, it has the advantage that since it is a perturbative method, it does not require the full explicit solution of the background geometry, but only its expansion around the boundary.

Finally, the investigation of the thermalization process in the boundary CFT, requires the study of black hole formation in the bulk \cite{AbajoArrastia:2010yt}, and, thus, the study of not static geometries. In such cases, the problem cannot be reduced to the problem of a co-dimension one minimal surface in a Riemannian manifold. Therefore, the geometric flow method that we presented has to be reformulated for co-dimension two minimal surfaces.

\subsection*{Acknowledgements}
The research of I.M. and G.P. has received funding from the Hellenic Foundation for Research and Innovation (HFRI) and the General Secretariat for Research and Technology (GSRT), in the framework of the ``First Post-doctoral researchers support'', under grant agreement No 2595. The research of D.K. is co-financed by Greece and the European Union (European Social Fund- ESF) through the Operational Programme ``Human Resources Development, Education and Lifelong Learning'' in the context of the project ``Strengthening Human Resources Research Potential via Doctorate Research'' (MIS-5000432), implemented by the State Scholarships Foundation (IKY). The authors would like to thank M. Axenides and E. Floratos for useful discussions.

%%%-----------------------------------------------------------------------------------------------------------------------------------------------------------------
\appendix

\renewcommand{\theequation}{\Alph{section}.\arabic{equation}}

\setcounter{equation}{0}
\section{The Embedding of the Minimal Surface in the Bulk}
\label{sec:flow_derivation}

In this appendix, we provide some intermediate steps in the derivation of the basic equation \eqref{eq:flow_Hd}, which describes a static minimal surface in an asymptotic AdS space as a geometric flow of the entangling surface towards the interior of the bulk. Since the defining property of the minimal surface is its vanishing mean curvature, we need to calculate the components of the second fundamental form, for the embedding of the minimal surface in the bulk, in the particular parametrization \eqref{eq:parametrization} that we use.

In the following the greek indices identify the coordinates in the bulk, including the holographic coordinate, thus they take $d$ distinct values. The latin indices $i$, $j$ and so on, identify the coordinates that parametrize a constant-$r$ plane in the bulk, thus, they take $d - 1$ distinct values. Finally, the latin indices $a$, $b$ and so on identify the variables that parametrize the intersection of the minimal surface with the constant-$r$ plane and thus, they take $d - 2$ distinct values.

Let us first derive some relations that are going to be useful in the following. The form of the parametrization of the minimal surface \eqref{eq:parametrization_bulk} and the particular choice of the parameters $u^a$ that satisfy \eqref{eq:parametrization} imply that
\begin{align}
{\partial _\rho } &= \frac{{\partial r}}{{\partial \rho }}{\partial _r} + \frac{{\partial {x^k}}}{{\partial \rho }}{\partial _k} = {\partial _r} + a{n^k}{\partial _k} , \label{eq:derivation_partial_rho}\\
{\partial _a} &= \frac{{\partial r}}{{\partial {u^a}}}{\partial _r} + \frac{{\partial {x^k}}}{{\partial {u^a}}}{\partial _k} = \frac{{\partial {x^k}}}{{\partial {u^a}}}{\partial _k} . \label{eq:derivation_partial_a}
\end{align}

Furthermore the parametrization \eqref{eq:parametrization} implies that
\begin{equation}
\frac{{\partial^2 {x^j}}}{{\partial {u^a} \partial {\rho}}} = {\partial _a}\left( {a{n^j}} \right) = \left( {{\partial _a}a} \right){n^j} + a{\partial _a}{n^j} .
\label{eq:derivation_mixed}
\end{equation}

The normal vector is normalized, i.e. ${n^i}{n^j}{h_{ij}} = 1$. This implies that
\begin{align}
2\left( {{\partial _\rho }{n^i}} \right){n^j}{h_{ij}} + {n^i}{n^j}{\partial _\rho }{h_{ij}} &= 0 ,\\
2\left( {{\partial _a}{n^i}} \right){n^j}{h_{ij}} + {n^i}{n^j}{\partial _a}{h_{ij}} &= 0 .
\end{align}
The above equations combined with the equations \eqref{eq:derivation_partial_rho} and \eqref{eq:derivation_partial_a} yield
\begin{align}
\left( {{\partial _\rho }{n^i}} \right){n^j}{h_{ij}} &=  - \frac{1}{2}{n^i}{n^j}\left( {{\partial _r}{h_{ij}} + a{n^k}{\partial _k}{h_{ij}}} \right) , \label{eq:derivation_drn_normal}\\
\left( {{\partial _a}{n^i}} \right){n^j}{h_{ij}} &=  - \frac{1}{2}{n^i}{n^j}\frac{{\partial {x^k}}}{{\partial {u^a}}}{\partial _k}{h_{ij}} . \label{eq:derivation_dan_normal}
\end{align}

The specific choice of the parameters $u^a$ \eqref{eq:parametrization} implies that ${n^i}\frac{{\partial {x^j}}}{{\partial {u^a}}}{h_{ij}} = 0$. It follows that
\begin{equation}
{\partial _\rho }{n^i}\frac{{\partial {x^j}}}{{\partial {u^a}}}{h_{ij}} + {n^i} \frac{{\partial^2 {x^j}}}{{\partial {u^a} \partial {\rho}}} {h_{ij}} + {n^i}\frac{{\partial {x^j}}}{{\partial {u^a}}}{\partial _\rho }{h_{ij}} = 0 .
\end{equation}
Implementing the equation \eqref{eq:derivation_mixed}, the above equation assumes the form
\begin{equation}
 - {\partial _\rho }{n^i}\frac{{\partial {x^j}}}{{\partial {u^a}}}{h_{ij}} = {\partial _a}a + a{n^i}{\partial _a}{n^j}{h_{ij}} + {n^i}\frac{{\partial {x^j}}}{{\partial {u^a}}}{\partial _\rho }{h_{ij}} .
\end{equation}
Finally, equations \eqref{eq:derivation_partial_rho} and \eqref{eq:derivation_dan_normal} allow the re-expression of the above equation as
\begin{equation}
- {\partial _\rho }{n^i}\frac{{\partial {x^j}}}{{\partial {u^a}}}{h_{ij}} = {\partial _a}a - \frac{1}{2}a{n^i}{n^j}\frac{{\partial {x^k}}}{{\partial {u^a}}}{\partial _k}{h_{ij}} + {n^i}\frac{{\partial {x^j}}}{{\partial {u^a}}}{\partial _r}{h_{ij}} + a\frac{{\partial {x^j}}}{{\partial {u^a}}}{n^i}{n^k}{\partial _k}{h_{ij}} .
\label{eq:derivation_drn_tangent}
\end{equation}

Let us now calculate the components of the second fundamental form for the embedding of the minimal surface in the bulk. We start with the $\rho\rho$ component. This equals
\begin{equation}
{K_{\rho\rho}} = - {\nabla _\kappa }{N^\mu }\frac{{\partial {x^\kappa }}}{{\partial \rho}}\frac{{\partial {x^\nu }}}{{\partial \rho}}{G_{\mu \nu }} ,
\end{equation}
where $G$ is the bulk metric that corresponds to the line element \eqref{eq:metric_G}. The indices $\mu$ and $\nu$ may be equal to $r$ or to any other value $i$. Since the bulk metric does not contain $ri$ elements, we get
\begin{equation}
{K_{\rho\rho}} =  - {\nabla _\kappa }{N^r}\frac{{\partial {x^\kappa }}}{{\partial \rho}}f - {\nabla _\kappa }{N^i}\frac{{\partial {x^\kappa }}}{{\partial \rho}}\frac{{\partial {x^j}}}{{\partial \rho}}{h_{ij}} .
\end{equation}
Then, implementing the definition of the covariant derivative $\nabla_\kappa$ in terms of the Christoffel symbols, we get
\begin{equation}
{K_{\rho\rho}} =  - {\partial _\rho}{N^r}f - \Gamma _{\kappa \lambda }^r{N^\lambda }\frac{{\partial {x^\kappa }}}{{\partial \rho}}f - {\partial _\rho}{N^i}\frac{{\partial {x^j}}}{{\partial \rho}}{h_{ij}} - \Gamma _{\kappa \lambda }^i{N^\lambda }\frac{{\partial {x^\kappa }}}{{\partial \rho}}\frac{{\partial {x^j}}}{{\partial \rho}}{h_{ij}} .
\end{equation}
The equation \eqref{eq:Christoffel_G} states that the Christoffel symbols with two $r$ indices vanish, hence,
\begin{multline}
{K_{\rho\rho}} =  - {\partial _\rho}{N^r}f - \Gamma _{rr}^r{N^r}f - \Gamma _{kl}^r{N^l}\frac{{\partial {x^k}}}{{\partial \rho}}f\\
- {\partial _\rho}{N^i}\frac{{\partial {x^j}}}{{\partial \rho}}{h_{ij}} - \Gamma _{kr}^i{N^r}\frac{{\partial {x^k}}}{{\partial \rho}}\frac{{\partial {x^j}}}{{\partial \rho}}{h_{ij}} - \Gamma _{rl}^i{N^l}\frac{{\partial {x^j}}}{{\partial \rho}}{h_{ij}} - \Gamma _{kl}^i{N^l}\frac{{\partial {x^k}}}{{\partial \rho}}\frac{{\partial {x^j}}}{{\partial \rho}}{h_{ij}} .
\end{multline}
We now take advantage of the particular parametrization \eqref{eq:parametrization}. In this parametrization, it holds that $N^i = c n^i$ and $N^r = - {ca}/{f}$. Furthermore, we substitute the values of the Christoffel symbols from equation \eqref{eq:Christoffel_G} and after some algebra we find
\begin{equation}
{K_{\rho\rho}} = \sqrt {f} c{\partial _\rho}\left( {\frac{a}{{\sqrt {f} }}} \right) - ca\left( {{\partial _\rho}{n^i}} \right){n^j}{h_{ij}} + \frac{{c{a^3}}}{{2f}}{n^k}{n^j}{\partial _r}{h_{jk}} - c{a^2}\gamma _{kl}^i{n^l}{n^k}{n^j}{h_{ij}} .
\end{equation}
At this point it is useful to implement the equation \eqref{eq:derivation_drn_normal}, which allows the re-expression of the above equation as
\begin{multline}
{K_{\rho\rho}} = \sqrt {f} c{\partial _\rho}\left( {\frac{a}{{\sqrt {f} }}} \right) + \frac{ca}{2}\left( {1 + \frac{{{a^2}}}{{f}}} \right){n^i}{n^j}{\partial _r}{h_{ij}} \\
+ \frac{c{a^2}}{2} \left( {\partial _k}{h_{ij}} - \gamma _{kj}^l {h_{il}} - \gamma _{ki}^l {h_{lj}} \right) {n^i}{n^j}{n^k} .
\end{multline}
The parentheses in the last term contain the covariant derivative of the metric $h_{ij}$ with respect to itself, thus it vanishes. Finally, using the fact that $c^{-2} = 1 + {{a^2}}/{{f}}$, we find
\begin{equation}
{K_{\rho\rho}} = \sqrt {f} c{\partial _\rho}\left( {\frac{a}{{\sqrt {f} }}} \right) + \frac{a}{{2c}}{n^i}{n^j}{\partial _r}{h_{ij}} .
\label{eq:derivation_Krr}
\end{equation}

We proceed to the $\rho a$ element of the second fundamental form. We recall that $\frac{\partial r}{\partial u^a} = 0$, $G_{ri} = 0$ and $\Gamma^i_{rr} = 0$. Then, ${K_{\rho a}}$ is given by
\begin{equation}
\begin{split}
{K_{\rho a}} &=  - {\nabla _\kappa }{N^\mu }\frac{{\partial {x^\kappa }}}{{\partial \rho}}\frac{{\partial {x^\nu }}}{{\partial {u^a}}}{G_{\mu \nu }} =  - {\nabla _\kappa }{N^i}\frac{{\partial {x^\kappa }}}{{\partial \rho}}\frac{{\partial {x^j}}}{{\partial {u^a}}}{h_{ij}}\\
 &=  - {\partial _\rho}{N^i}\frac{{\partial {x^j}}}{{\partial {u^a}}}{h_{ij}} - \Gamma _{\kappa \lambda }^i{N^\lambda }\frac{{\partial {x^\kappa }}}{{\partial \rho}}\frac{{\partial {x^j}}}{{\partial {u^a}}}{h_{ij}}\\
 &=  - {\partial _\rho}{N^i}\frac{{\partial {x^j}}}{{\partial {u^a}}}{h_{ij}} - \Gamma _{rl}^i{N^l}\frac{{\partial {x^j}}}{{\partial {u^a}}}{h_{ij}} - \Gamma _{kr}^i{N^r}\frac{{\partial {x^k}}}{{\partial \rho}}\frac{{\partial {x^j}}}{{\partial {u^a}}}{h_{ij}} - \Gamma _{kl}^i{N^l}\frac{{\partial {x^k}}}{{\partial \rho}}\frac{{\partial {x^j}}}{{\partial {u^a}}}{h_{ij}}.
\end{split}
\end{equation}
Finally, substituting the values of the Christoffel symbols from the equation \eqref{eq:Christoffel_G} and the components of the vector $N$ in terms of components of the vector $n$ and the functions $c$ and $a$, as we did for the $K_{\rho \rho}$ component, we find
\begin{equation}
{K_{\rho a}} = - c {\partial _\rho}{n^i}\frac{{\partial {x^j}}}{{\partial {u^a}}}{h_{ij}} - \frac{c}{2}{n^l}\frac{{\partial {x^j}}}{{\partial {u^a}}}{\partial _r}{h_{jl}} + \frac{c a^2}{2 f}{{n^k}}\frac{{\partial {x^j}}}{{\partial {u^a}}}{\partial _r}{h_{jk}} - c a \gamma _{kl}^i{n^l}{{n^k}}\frac{{\partial {x^j}}}{{\partial {u^a}}}{h_{ij}} .
\end{equation}
Implementation of the equation \eqref{eq:derivation_drn_tangent} yields
\begin{multline}
{K_{\rho a}} = c{\partial _a}a - \frac{1}{2}ca{n^i}{n^j}\frac{{\partial {x^k}}}{{\partial {u^a}}}{\partial _k}{h_{ij}} + ca{n^i}{n^k}\frac{{\partial {x^j}}}{{\partial {u^a}}}{\partial _k}{h_{ij}} \\
+ \frac{c}{2}\left( {1 + \frac{{{a^2}}}{{f}}} \right){n^i}\frac{{\partial {x^j}}}{{\partial {u^a}}}{\partial _r}{h_{ij}} - ca\gamma _{kl}^i{n^l}{n^k}\frac{{\partial {x^j}}}{{\partial {u^a}}}{h_{ij}} .
\end{multline}
Using the fact that $c^{-2} = 1 + {{a^2}}/{{f}}$ and after an appropriate relabelling of some indices we find
\begin{multline}
{K_{\rho a}} = c{\partial _a}a + \frac{1}{{2c}}{n^i}\frac{{\partial {x^j}}}{{\partial {u^a}}}{\partial _r}{h_{ij}} \\
+ ca{n^i} \left( {n^k}\frac{{\partial {x^j}}}{{\partial {u^a}}} - \frac{1}{2}{n^j}\frac{{\partial {x^k}}}{{\partial {u^a}}} \right) \left( {{\partial _k}{h_{ij}} - \gamma _{ki}^l{h_{lj}} - \gamma _{kj}^l{h_{il}}} \right) .
\end{multline}
The last parentheses contain the covariant derivative of the metric $h_{ij}$ with respect to itself, therefore it vanishes. So we are left with
\begin{equation}
{K_{\rho a}} = c{\partial _a}a + \frac{1}{{2c}}{n^i}\frac{{\partial {x^j}}}{{\partial {u^a}}}{\partial _r}{h_{ij}} .
\end{equation}

The $ab$ element of the second fundamental form for the embedding of the minimal surface in the bulk is given by
\begin{equation}
{K_{ab}} =  - {\nabla _\kappa }{N^\mu }\frac{{\partial {x^\kappa }}}{{\partial {u^a}}}\frac{{\partial {x^\nu }}}{{\partial {u^b}}}{G_{\mu \nu }} .
\end{equation}
If either $\kappa$ or $\nu$ is equal to $r$ the partial derivatives are vanishing. Thus, the above expression can be simplified to
\begin{equation}
{K_{ab}} = - {\nabla _k}{N^i}\frac{{\partial {x^k}}}{{\partial {u^a}}}\frac{{\partial {x^j}}}{{\partial {u^b}}}{h_{ij}}.
\end{equation}
We write the covariant derivative in terms of the Christoffel symbols to find
\begin{equation}
\begin{split}
{K_{ab}} &=  - {\partial _k}{N^i}\frac{{\partial {x^k}}}{{\partial {u^a}}}\frac{{\partial {x^j}}}{{\partial {u^b}}}{h_{ij}} - \Gamma _{k\lambda }^i{N^\lambda }\frac{{\partial {x^k}}}{{\partial {u^a}}}\frac{{\partial {x^j}}}{{\partial {u^b}}}{h_{ij}}\\
 &=  - {\partial _a}{N^i}\frac{{\partial {x^j}}}{{\partial {u^b}}}{h_{ij}} - \Gamma _{kr}^i{N^r}\frac{{\partial {x^k}}}{{\partial {u^a}}}\frac{{\partial {x^j}}}{{\partial {u^b}}}{h_{ij}} - \Gamma _{kl}^i{N^l}\frac{{\partial {x^k}}}{{\partial {u^a}}}\frac{{\partial {x^j}}}{{\partial {u^b}}}{h_{ij}} .
\end{split}
\end{equation}
We substitute the Christoffel symbols from equation \eqref{eq:Christoffel_G}, as well as $N^i = c n^i$ and $N^r = - {ca}/{f}$, and we find
\begin{equation}
{K_{ab}} =  - c{\partial _a}{n^i}\frac{{\partial {x^j}}}{{\partial {u^b}}}{h_{ij}} + \frac{{ca}}{{2 f}}\frac{{\partial {x^k}}}{{\partial {u^a}}}\frac{{\partial {x^j}}}{{\partial {u^b}}} {\partial _r}{h_{kj}} - \gamma _{kl}^ic{n^l}\frac{{\partial {x^k}}}{{\partial {u^a}}}\frac{{\partial {x^j}}}{{\partial {u^b}}}{h_{ij}} .
\end{equation}
Taking into account the equation \eqref{eq:embedding_k}, we have
\begin{equation}
{K_{ab}} = c{k_{ab}} + \frac{{ca}}{{2 f}} \frac{{\partial {x^k}}}{{\partial {u^a}}}\frac{{\partial {x^j}}}{{\partial {u^b}}} {\partial _r}{h_{kj}} .
\label{eq:derivation_Kab}
\end{equation}

It is now simple to calculate the trace of the second fundamental form, using equations \eqref{eq:embedding_inverse_G}, \eqref{eq:derivation_Krr} and \eqref{eq:derivation_Kab},
\begin{equation}
\begin{split}
K &= {\Gamma ^{\rho\rho}}{K_{\rho\rho}} + {\Gamma ^{ab}}{K_{ab}}\\
&= ck + \frac{{c^3}}{{\sqrt {f} }}{\partial _\rho}\left( {\frac{a}{{\sqrt {f} }}} \right) + \frac{{ca}}{{2 f}}\left( {{\gamma ^{ab}}\frac{{\partial {x^i}}}{{\partial {u^a}}}\frac{{\partial {x^j}}}{{\partial {u^b}}} + {n^i}{n^j}} \right) {\partial _r}{h_{ij}}\\
&= ck + \frac{{c^3}}{{\sqrt {f} }}{\partial _\rho}\left( {\frac{a}{{\sqrt {f} }}} \right) + \frac{{ca}}{{2 f}}{h^{ij}}{{\partial _r}{h_{ij}}} .
\end{split}
\end{equation}

\setcounter{equation}{0}
\section{A Non-trivial Verifying Solution of the Flow Equation}
\label{sec:simple solutions}

It is quite trivial to show that several explicitly known minimal surfaces, which possess either rotational or translational symmetry, satisfy the equation \eqref{eq:flow_Hd}. These include the minimal surfaces that correspond to a spherical entangling surface or a strip region in AdS$_{d+1}$ and the catenoid minimal surfaces in AdS$_4$. In all these cases, the symmetry allows the reduction of \eqref{eq:flow_Hd} to an ordinary differential equation for a single variable. As a non-trivial verifying example, we will study the case of a helicoid minimal surface in AdS$_4$ in \Poincare coordinates. In this case the boundary data depend on the position on the entangling curve and the equation \eqref{eq:flow_Hd} is a non-trivial partial differential equation.

The equation of the helicoid \cite{Pastras:2016vqu} is
\begin{equation}
r = \sqrt{e^{- 2\omega \phi} - x^2} .
\end{equation}
We will use the following parametrization
\begin{equation}
\begin{split}
r &= \rho,\\
\phi &= \phi \left( \rho, u \right) ,\\
x &= \sqrt{e^{- 2\omega \phi \left( \rho, u \right)} - \rho^2}
\end{split}
\label{eq:simple_catenoid_parametrization}
\end{equation}
and specify the function $\phi \left( \rho, u \right)$ so that the parametrization obeys the equation \eqref{eq:parametrization}. This is equivalent to imposing $\Gamma_{u\rho} = 0$, i.e.,
\begin{equation}\label{eq:hel_diagonal}
\partial_u x \partial_z x + x^2 \partial_u \phi \partial_z \phi = 0.
\end{equation}
Substituting \eqref{eq:simple_catenoid_parametrization} in \eqref{eq:hel_diagonal} yields
\begin{equation}
 e^{2\omega \phi}\partial_\rho \phi \left[ \left( e^{-2\omega \phi} - \rho^2 \right)^2 + \omega^2 e^{-4\omega \phi} \right] + \omega \rho = 0.
\end{equation}
This equation has the solution
\begin{align}
2{e^{ - 2\omega \varphi }} = u\left( {1 + {\omega ^2}} \right) + {\rho ^2} + \sqrt {{u^2}{{\left( {1 + {\omega ^2}} \right)}^2} + 2u\left( {{\omega ^2} - 1} \right){\rho ^2} + {\rho ^4}} , \label{eq:simple_helicoid_phi} \\
2{x^2} = u\left( {1 + {\omega ^2}} \right) - {\rho ^2} + \sqrt {{u^2}{{\left( {1 + {\omega ^2}} \right)}^2} + 2u\left( {{\omega ^2} - 1} \right){\rho ^2} + {\rho ^4}} . \label{eq:simple_helicoid_x}
\end{align}

In the special parametrization \eqref{eq:parametrization}, it holds that $\partial_\rho x = a n^x$, $\partial_\rho \varphi = a n^\varphi$. Thus, the normalization of the vector $n^i$ reads
\begin{equation}
a = \frac{1}{\rho }{\left[ {{{\left( {{\partial _\rho }x} \right)}^2} + {x^2}{{\left( {{\partial _\rho }\varphi } \right)}^2}} \right]^{\frac{1}{2}}} .
\end{equation}
Substituting the equations \eqref{eq:simple_helicoid_phi} and \eqref{eq:simple_helicoid_x} yields
\begin{equation}
{\left( {a\rho } \right)^2} = \frac{{u\left( {1 + {\omega ^2}} \right) + {\rho ^2} - \sqrt {{u^2}{{\left( {1 + {\omega ^2}} \right)}^2} + 2u\left( {{\omega ^2} - 1} \right){\rho ^2} + {\rho ^4}} }}{{2\sqrt {{u^2}{{\left( {1 + {\omega ^2}} \right)}^2} + 2u\left( {{\omega ^2} - 1} \right){\rho ^2} + {\rho ^4}} }} .
\label{eq:simple_helicoid_arho}
\end{equation}
 
On the constant-$r$ plane the metric reads ${ds}^2 = \frac{1}{\rho^2} \left( {dx}^2 + x^2 {d\varphi}^2 \right)$. Thus, the non-vanishing Christoffel symbols are $\gamma^x_{\varphi\varphi} = - x$, $\gamma^x_{x\varphi} = 1 / x$. Thus, using its definition, the second fundamental form equals
\begin{equation}
\begin{split}
{\rho ^2}{k_{uu}} &=  - \left( {{\partial _u}{n^x}} \right)\left( {{\partial _u}x} \right) - {x^2}\left( {{\partial _u}{n^\varphi }} \right)\left( {{\partial _u}\varphi } \right) - x{n^x}{\left( {{\partial _u}\varphi } \right)^2}\\
&=  - \frac{1}{a}\left[ {\left( {{\partial _u}x} \right)\left( {{\partial _\rho }{\partial _u}x} \right) - {x^2}\left( {{\partial _u}\varphi } \right)\left( {{\partial _\rho }{\partial _u}\varphi } \right) - x\left( {{\partial _u}x} \right){{\left( {{\partial _u}\varphi } \right)}^2}} \right] =  - \frac{{{\partial _\rho }\left( {{\rho ^2}{\gamma _{uu}}} \right)}}{{2a}} ,
\end{split}
 \end{equation}
since
\begin{equation}
{\rho ^2}{\gamma _{uu}} = {\left( {{\partial _u}x} \right)^2} - {x^2}{\left( {{\partial _u}\varphi } \right)^2} .
\label{eq:simple_helicoid_gamma}
\end{equation}
The intersection of the minimal surface with the constant-$r$ plane is in this case one-dimensional. Thus, trivially, ${\gamma ^{uu}} = 1/{\gamma _{uu}}$ and
\begin{equation}
2ka =  - \frac{{{\partial _\rho }\left( {{\rho ^2}{\gamma _{uu}}} \right)}}{{{\rho ^2}{\gamma _{uu}}}} .
\label{eq:simple_helicoid_ka}
\end{equation}
Finally, upon substitution of \eqref{eq:simple_helicoid_phi} and \eqref{eq:simple_helicoid_x} in \eqref{eq:simple_helicoid_gamma}, we find
\begin{multline}
{\rho ^2}{\gamma _{uu}} = \frac{1}{8}{\left[ {\left( {1 + {\omega ^2}} \right) + \frac{{u\left( {1 + {\omega ^2}} \right) + \left( {{\omega ^2} - 1} \right){\rho ^2}}}{{\sqrt {{u^2}{{\left( {1 + {\omega ^2}} \right)}^2} + 2u\left( {{\omega ^2} - 1} \right){\rho ^2} + {\rho ^4}} }}} \right]^2}\\
 \times \frac{{u\left( {1 + {\omega ^2}} \right) + \sqrt {{u^2}{{\left( {1 + {\omega ^2}} \right)}^2} + 2u\left( {{\omega ^2} - 1} \right){\rho ^2} + {\rho ^4}} }}{{u\left( {1 + {\omega ^2}} \right) + {\rho ^2} + \sqrt {{u^2}{{\left( {1 + {\omega ^2}} \right)}^2} + 2u\left( {{\omega ^2} - 1} \right){\rho ^2} + {\rho ^4}} }} .
\label{eq:simple_helicoid_gamma2}
\end{multline}

It is now a matter of tedious algebra to show that upon substitution of \eqref{eq:simple_helicoid_arho}, \eqref{eq:simple_helicoid_ka} and \eqref{eq:simple_helicoid_gamma2} into \eqref{eq:flow_Hd}, the latter is satisfied.

\setcounter{equation}{0}
\section{The Divergent Terms of Entanglement Entropy for Spherical Entangling Surfaces}
\label{sec:terms_simple}

In this appendix, we calculate all the divergent terms of the expansion of the entanglement entropy in the case of a spherical entangling surface in AdS$_{d+1}$, taking advantage of the fact that the minimal surface is explicitly known, in order to compare with the general formulae of section \ref{sec:entropy}.

We adopt polar coordinates on the constant-$r$ plane. Let $x$ denote the radial coordinate, i.e. $x = \sqrt{x^i x^i}$. Then the bulk metric assumes the form
\begin{equation}
{ds}^2 = \frac{1}{{{r^2}}}\left( {{dr}^2 - d{t^2} + {dx}^2 + {x^2}d\Omega _{d - 2}^2} \right) .
\end{equation}
The minimal surface, corresponding to a spherical entangling surface of radius $R$ is given by
\begin{equation}
r \left( x \right) = \sqrt {{R^2} - {x^2}} .
\end{equation}

We parametrize the minimal surface using $x$ and the $d - 2$ spherical coordinates on the constant-$r$ slices (and constant time slices). Then, the only non-trivial element of the induced metric for the embedding of the minimal surface in the bulk is
\begin{equation}
{\Gamma _{xx}} = \frac{1}{{r{{\left( x \right)}^2}}}\left( {1 + {{\left( {\frac{{dr\left( x \right)}}{{dx}}} \right)}^2}} \right) = \frac{{{R^2}}}{{{{\left( {{R^2} - {x^2}} \right)}^2}}} ,
\end{equation}
while all the others are directly inherited from the bulk metric, since the angular coordinates do not appear in the minimal surface equation. Thus, the induced metric on the minimal surface is given by
\begin{equation}
d{s^2} = \frac{1}{{{R^2} - {x^2}}}\left( {\frac{{{R^2}}}{{{R^2} - {x^2}}}d{x^2} + {x^2}d\Omega _{d - 2}^2} \right) .
\end{equation}
The area element of the minimal surface can thus be expressed as
\begin{equation}
dA = \frac{{R{x^{d - 2}}}}{{{{\left( {{R^2} - {x^2}} \right)}^{\frac{d}{2}}}}}dx d{\Omega _{d - 2}} .
\end{equation}

We cutoff the minimal surface at $r = 1 / \Lambda$ . This is equivalent to restricting to the region $x < \sqrt {{R^2} - 1 / {\Lambda^2}} $. Thus, the area of the cut-off minimal surface equals
\begin{equation}
\begin{split}
A\left( d ; {\Lambda} \right) &= \int {d{\Omega _{d - 2}}} \int_0^{\sqrt {{R^2} - 1 / {\Lambda^2}}} {\frac{{R{x^{d - 2}}}}{{{{\left( {{R^2} - {x^2}} \right)}^{\frac{d}{2}}}}}dx} \\
&= \frac{{\mathcal{A}_{d - 2}}}{{R^{d - 2}}} {{\rm B}_{1 - \frac{{{1}}}{{{R^2 \Lambda^2}}}}}\left( {\frac{{d - 1}}{2} , - \frac{{d - 2}}{2}} \right) ,
\end{split}
\end{equation}
where $\mathcal{A}_d$ is the area of a $d$-dimensional sphere with radius $R$ (thus $\mathcal{A}_{d - 2}$ is the area of the entangling surface) and ${\rm B}_x\left(a,b\right)$ is the incomplete beta function.

For $d = 2,3,4,5$, the above expression reads
\begin{align}
A\left( {2;\Lambda} \right) &= 2{\tanh ^{ - 1}}\sqrt {1 - \frac{1}{R^2 \Lambda^2}} , \\
A\left( {3;\Lambda} \right) &= 2\pi \left( {{R}{\Lambda} - 1} \right) , \\
A\left( {4;\Lambda} \right) &= 2\pi \left( {{R^2}{\Lambda^2}\sqrt {1 - \frac{1}{R^2 \Lambda^2}}  - {{\tanh }^{ - 1}}\sqrt {1 - \frac{1}{R^2 \Lambda^2}} } \right) , \\
A\left( {5;\Lambda} \right) &= \frac{{2{\pi ^2}}}{3}\left( {{R^3}{\Lambda^3} - 3{R}{\Lambda} + 2} \right) .
\end{align}

It is possible to derive explicit formulae at all dimensions using the recursive relation
\begin{equation}
b{{\rm B}_x}\left( {a,b} \right) = \left( {a - 1} \right){{\rm B}_x}\left( {a - 1,b + 1} \right) - {x^{a - 1}}{\left( {1 - x} \right)^b} .
\end{equation}
We also recall that
\begin{equation}
\mathcal{A}_d = 2 \frac{{{\pi ^{\frac{{d + 1}}{2}}}}}{{\Gamma \left( {\frac{{d + 1}}{2}} \right)}} R^d .
\end{equation}
The above imply that
\begin{equation}
\begin{split}
A\left( {d + 2;\Lambda} \right) &= \frac{{{\pi ^{\frac{{d + 1}}{2}}}}}{{\Gamma \left( {\frac{{d + 1}}{2}} \right)}}{B_{1 - \frac{1}{R^2 \Lambda^2}}}\left( {\frac{{d + 1}}{2}, - \frac{d}{2}} \right)\\
 &=  - \frac{{4\pi }}{{d\left( {d - 1} \right)}}\frac{{{\pi ^{\frac{{d - 1}}{2}}}}}{{\Gamma \left( {\frac{{d - 1}}{2}} \right)}}\left[ {\frac{{d - 1}}{2}{B_{1 - \frac{1}{R^2 \Lambda^2}}}\left( {\frac{{d - 1}}{2}, - \frac{{d - 2}}{2}} \right) }^{\phantom{\frac{d}{2}}} \right.\\
 &\quad\quad\quad\quad\quad\quad\quad\quad\quad\quad\quad\quad \left.{ - {{\left( {1 - \frac{1}{R^2 \Lambda^2}} \right)}^{\frac{{d - 1}}{2}}}{{\left( {{R^2}{\Lambda^2}} \right)}^{\frac{d}{2}}}} \right]\\
 &=  - \frac{{2\pi }}{d}A\left( {d;\Lambda} \right) + \frac{2}{d}\frac{{{\pi ^{\frac{{d + 1}}{2}}}}}{{\Gamma \left( {\frac{{d + 1}}{2}} \right)}}{R}{\Lambda}{\left( {{R^2}{\Lambda^2} - 1} \right)^{\frac{{d - 1}}{2}}} .
\end{split}
\label{eq:sphere_iterative}
\end{equation}

For odd $d = 2 k + 1$, the above formula can be written as
\begin{equation}
A\left( {2k + 1;\Lambda} \right) =  - \frac{{2\pi }}{{2k - 1}}A\left( {2k - 1;\Lambda} \right) + \frac{2}{{2k - 1}}\frac{{{\pi ^d}}}{{\left( {k - 1} \right)!}}{R}{\Lambda}{\left( {{R^2}{\Lambda^2} - 1} \right)^{k - 1}} .
\end{equation}
This equation, combined with the fact that $A \left( {1 ; \Lambda} \right) = 1$, iteratively results in
\begin{equation}
A\left( {2k + 1;\Lambda} \right) = \frac{{{{\left( { - 2\pi } \right)}^k}}}{{\left( {2k - 1} \right)!!}}\left[ {1 - \sum\limits_{n = 0}^{k - 1} {\frac{{{{\left( { - 1} \right)}^n}\left( {2n - 1} \right)!!}}{{\left( {2n} \right)!!}}{R}{\Lambda}{{\left( {{R^2}{\Lambda^2} - 1} \right)}^n}} } \right] .
\end{equation}
The above is clearly a polynomial of ${R}{\Lambda}$ of order $2k - 1 = d - 2$, containing only odd powers of ${R}{\Lambda}$, except for a constant term. We can use Newton's binomial theorem in order to acquire an explicit form of this polynomial
\begin{equation}
\begin{split}
A\left( {2k + 1;\Lambda} \right) &= \frac{{{{\left( { - 2\pi } \right)}^k}}}{{\left( {2k - 1} \right)!!}}\left[ {1 - \sum\limits_{n = 0}^{k - 1} {\sum\limits_{m = 0}^n {\frac{{{{\left( { - 1} \right)}^m}\left( {2n - 1} \right)!!}}{{\left( {2n} \right)!!}}\frac{{n!}}{{m!\left( {n - m} \right)!}}{{\left( {{R}{\Lambda}} \right)}^{2m + 1}}} } } \right]\\
&= \frac{{{{\left( { - 2\pi } \right)}^k}}}{{\left( {2k - 1} \right)!!}}\left[ {1 - \sum\limits_{m = 0}^{k - 1} {\left[ \sum\limits_{n = m}^{k - 1} {\frac{{\left( {2n - 1} \right)!!}}{{2^n\left( {n - m} \right)!}}} \right] \frac{{{{\left( { - 1} \right)}^m}}}{{m!}} {{\left( {{R}{\Lambda}} \right)}^{2m + 1}} } } \right]\\
&= {\left( { - \pi } \right)^k}\left[ {\frac{{{2^k}}}{{\left( {2k - 1} \right)!!}} - 2\sum\limits_{m = 0}^{k - 1} {\frac{{{{\left( { - 1} \right)}^m}}}{{\left( {1 + 2m} \right)m!\left( {k - m - 1} \right)!}}{{\left( {{R}{\Lambda}} \right)}^{2m + 1}}} } \right] .
\end{split}
\end{equation}

Adopting the notation \eqref{eq:area_notation} we find that
\begin{equation}
\begin{split}
{a_{d - 2 - 2n}} &= \frac{{{{\left( {2\pi } \right)}^{\frac{{d - 1}}{2}}}}}{{{{\left( { - 2} \right)}^n}n!\left( {d - 2 - 2n} \right)\left( {d - 3 - 2n} \right)!!}}{R^{d - 2 - 2n}} \\
&= \frac{{\left( {d - 3} \right)!!}}{{{{\left( { - 2} \right)}^n}n!\left( {d - 2 - 2n} \right)\left( {d - 3 - 2n} \right)!!}}\frac{{{\mathcal{A}_{d - 2}}}}{{{R^{2n}}}} .
\end{split}
\label{eq:sphere_an}
\end{equation}

For completeness, we note that the constant finite term $\tilde{a}$ equals
\begin{equation}
\tilde{a} = \frac{{{{\left( { - 2\pi } \right)}^{\frac{{d - 1}}{2}}}}}{{\left( {d - 2} \right)!!}} = \frac{{{{\left( { - 1} \right)}^{\frac{{d - 1}}{2}}}2\left( {d - 3} \right)!!}}{{\left( {d - 2} \right)!!}} \frac{{{\mathcal{A}_{d - 2}}}}{{{R^{d - 2}}}}.
\end{equation}

For even $d = 2 k$ the iterative formula \eqref{eq:sphere_iterative} assumes the form
\begin{equation}
A\left( {2k;\Lambda} \right) =  - \frac{{2\pi }}{{2\left( {k - 1} \right)}}A\left( {2k - 2;\Lambda} \right) + \frac{2}{{2\left( {k - 1} \right)}}\frac{{{{\left( {2\pi } \right)}^{k - 1}}}}{{\left( {2k - 3} \right)!!}}{R}{\Lambda}{\left( {{R^2}{\Lambda^2} - 1} \right)^{k - \frac{3}{2}}} ,
\end{equation}
which combined with the fact that $A\left( {2;\Lambda} \right) = 2{\tanh ^{ - 1}}\sqrt {1 - \frac{1}{R^2 \Lambda^2}} $ results in
\begin{multline}
A\left( {2k;\Lambda} \right) = \frac{{2 {{\left( { - \pi } \right)}^{k - 1}}}}{{\left( {k - 1} \right)!}}\left[ {{{\tanh }^{ - 1}}\sqrt {1 - \frac{1}{R^2 \Lambda^2}} }\right.\\
\left. { - \sum\limits_{n = 0}^{k - 2} {\frac{{{{\left( { - 1} \right)}^n}\left( {2n} \right)!!}}{{\left( {2n + 1} \right)!!}}{{\left( {{R}{\Lambda}} \right)}^{n + 2}}{{\left( {1 - \frac{1}{R^2 \Lambda^2}} \right)}^{n + \frac{1}{2}}} } } \right] .
\end{multline}
If one expands the square root and the inverse hyperbolic tangent in powers of ${R}{\Lambda}$, it is evident that only even powers will appear, apart from a logarithmic term from the expansion of the inverse hyperbolic tangent. The polynomially divergent terms, which are denoted by $A^+ \left( {2k;\Lambda} \right)$, can be easily found, via the Taylor expansion of ${{\left( {1 - x} \right)}^{n + \frac{1}{2}}}$,
\begin{equation}
{\left( {1 - x} \right)^{n + \frac{1}{2}}} = \sum\limits_{m = 0}^\infty  {\frac{{{{\left( { - 1} \right)}^m}\left( {2n + 1} \right)!!}}{{m!{2^m}\left( {2n + 1 - 2m} \right)!!}}} {x^m} .
\end{equation}
Thus,
\begin{equation}
\begin{split}
{A^ + }\left( {2k;\Lambda } \right) &=  - 2\frac{{{{\left( { - \pi } \right)}^{k - 1}}}}{{\left( {k - 1} \right)!}}\sum\limits_{n = 0}^{k - 2} {\sum\limits_{m = 0}^n {\frac{{{{\left( { - 1} \right)}^{m + n}}\left( {2n} \right)!!}}{{m!{2^m}\left( {2n + 1 - 2m} \right)!!}}} {{\left( {{R}{\Lambda }} \right)}^{2n + 2 - 2m}}} \\
 &=  - 2\frac{{{{\left( { - \pi } \right)}^{k - 1}}}}{{\left( {k - 1} \right)!}}\sum\limits_{m = 0}^{k - 2} {\left[ {\sum\limits_{n = m}^{k - 2} {\frac{{n!}}{{\left( {n - m} \right)!}}} } \right]\frac{{{{\left( { - 2} \right)}^m}}}{{\left( {2m + 1} \right)!!}}{{\left( {{R}{\Lambda }} \right)}^{2m + 2}}} \\
 &=  - 2{\left( { - \pi } \right)^{k - 1}}\sum\limits_{m = 0}^{k - 2} {\frac{{{{\left( { - 2} \right)}^m}}}{{\left( {m + 1} \right)\left( {k - m - 2} \right)!\left( {2m + 1} \right)!!}}{{\left( {{R}{\Lambda }} \right)}^{2m + 2}}} .
\end{split}
\end{equation}
Adopting the same notation \eqref{eq:area_notation}, as in the case of odd $d$, it is clear that
\begin{equation}
\begin{split}
{a_{d - 2 - 2n}} &= 2{\left( {2\pi } \right)^{\frac{{d - 2}}{2}}}\frac{1}{{{{\left( { - 2} \right)}^n}\left( {d - 2 - 2n} \right)n!\left( {d - 3 - 2n} \right)!!}}{R^{d - 2 - 2m}}\\
 &= \frac{{\left( {d - 3} \right)!!}}{{{{\left( { - 2} \right)}^n}\left( {d - 2 - 2n} \right)n!\left( {d - 3 - 2n} \right)!!}}\frac{{{\mathcal{A}_{d - 2}}}}{{{R^{2n}}}} .
\end{split}
\end{equation}
Comparing to the equation \eqref{eq:sphere_an} we see that when expressed in terms of the area of the entangling surface, the coefficients $a_n$ are given by the same formula for both odd and even dimensions.

The logarithmic term emerges from the expansion ${{{\tanh }^{ - 1}}\sqrt {1 - x^2}} = - \ln x + \mathcal{O} \left( 1 \right)$. It follows that
\begin{equation}
{a_0} = 2\frac{{{{\left( { - \pi } \right)}^{k - 1}}}}{{\left( {k - 1} \right)!}} = 2\frac{{{{\left( { - 2\pi } \right)}^{\frac{{d - 2}}{2}}}}}{{\left( {d - 2} \right)!!}} = {\left( { - 1} \right)^{\frac{d - 2}{2}}} \frac{{\left( {d - 3} \right)!!}}{{\left( {d - 2} \right)!!}}{\mathcal{A}_{d - 2}}.
\label{eq:sphere_a0}
\end{equation}

Studying equation \eqref{eq:sphere_an}, we observe that the leading divergent terms are
\begin{equation}
\begin{split}
{a_{d - 2}} &= \frac{1}{{\left( {d - 2} \right)}}{\mathcal{A}_{d - 2}} ,\\
{a_{d - 4}} &=  - \frac{{d - 3}}{{2\left( {d - 4} \right)}}\frac{{{\mathcal{A}_{d - 2}}}}{{{R^2}}} ,\\
{a_{d - 6}} &= \frac{{\left( {d - 3} \right)\left( {d - 5} \right)}}{{8\left( {d - 6} \right)}}\frac{{{\mathcal{A}_{d - 2}}}}{{{R^4}}} .
\end{split}
\label{eq:sphere_a_d}
\end{equation}
The first one is the usual ``area law term''.

\end{document}